\documentclass[usenatbib,10pt,a4paper]{mnras} 
\usepackage[utf8]{inputenc}
\usepackage[T1]{fontenc}
\usepackage{aas_macros}
\usepackage{amssymb,amsmath}
\usepackage{booktabs} 
\usepackage{graphicx}
\usepackage{enumerate}
\usepackage{lmodern}
\usepackage{siunitx}
\usepackage{subcaption}
\usepackage{tikz}
\usepackage{xspace} 
  
\defcitealias{2015MNRAS.451..527L}{Paper I} 

\newcommand{\Mn}{\ensuremath{{M}_\mathrm{N}}}
\newcommand{\Mt}{\ensuremath{{M}_\mathrm{T}}}
\newcommand{\Msun}{\ensuremath{\mathrm{M}_\odot}}

\newcommand{\eps}{\ensuremath{\varepsilon}}
\newcommand{\mean}[1]{\ensuremath{\left\langle #1 \right\rangle}}
\newcommand{\abs}[1]{\ensuremath{\left\lvert#1\right\rvert}}

\newcommand{\vect}[1] {\ensuremath{\bmath{#1}}}
\newcommand{\tens}[1] {\ensuremath{\mathbfss{#1}}} %

\newcommand{\hGpc}[1]{\ensuremath{{#1}\,h^{-1} \mathrm{Gpc}}}

\newcommand{\hMsun}[1]{\ensuremath{\num{#1}\,h^{-1} \Msun}}

\newcommand{\rvir}{\ensuremath{R_\text{vir}}}

\begin{document}

\title[Ecology of haloes II]{Ecology of dark matter haloes --II.
  Effects of interactions on the alignment of halo pairs
}
\author[B.~L'Huillier, C.~Park and J.~Kim]
{Benjamin~L'Huillier,$^{1,2}$
  Changbom~Park$^1$
  and 
  Juhan~Kim$^{3}$\thanks{E-mail:benjamin@kasi.re.kr      (BL),
    cbp@kias.re.kr (CBP), kjhan@kias.re.kr (JHK)
  } \\
  $^1$ School of Physics, Korea Institute for Advanced
  Study, 85 Hoegi-ro, Dongdaemun-gu, Seoul 130-722, Korea\\
  $^2$ Korea Astronomy and Space Science Institute, 776 Daedeok daero,
  Yuseong-gu, Daejeon, Korea\\ 
  $^3$ Center for Advanced Computation, Korea Institute for Advanced 
  Study, 85 Hoegi-ro, Dongdaemun-gu, Seoul 130-722, Korea
} 

\date{Accepted 2017 January 16. Received 2017 January 13; in original form 2016 March 20}

\pagerange{\pageref{firstpage}--\pageref{lastpage}} \pubyear{2016}

\maketitle
 
\label{firstpage}

\begin{abstract}
  We use the Horizon Run  4 cosmological $N$-body simulation to study
  the effects of  distant and close interactions on  the alignments of
  the  shapes,  spins,  and  orbits   of  targets  haloes  with  their
  neighbours, and their dependence on the local density environment and
  neighbour separation.
  Interacting  targets  have a  significantly  lower  spin and  higher
  sphericity and oblateness than all targets.
  Interacting pairs initially have  anti-parallel spins, but the spins
  develop parallel alignment as time goes on.
  Neighbours tend  to evolve in the  plane of rotation of  the target,
  and in the direction of the major axis of prolate haloes.
  Moreover, interactions  are preferentially radial, while  pairs with
  non-radial orbits are preferentially prograde.
  The alignment signals are stronger at high-mass and for
  close separations, and independent on the large-scale density.
  Positive alignment signals are found at redshifts up to 4, and
  increase with decreasing redshifts.
  Moreover, the orbits tend to  become prograde at low redshift, while
  no alignment is found at high redshift ($z=4$).  
\end{abstract}

 \begin{keywords}
   {Methods: numerical -- Galaxies: haloes, interactions -- Cosmology:
     Large-scale structure of the Universe, Theory, Dark matter} 
 \end{keywords}

\section{Introduction} 

Galaxy  surveys and  cosmological simulations  have revealed  that the
Universe  is  structured  on  large-scale  as a  cosmic  web  made  of
two-dimensional  walls  and   one-dimensional  filaments  encompassing
voids, with clusters in the knots \citep{1986ApJ...302L...1D}.  
This cosmic web is the natural outcome of the gravitational collapse of
the primordial density fluctuations \citep{1996Natur.380..603B}.  
Galaxies  evolve in  this  cosmic web,  accreting  material along  the
filaments. 
The anisotropic  nature of the cosmic  web is thus expected  to affect
the development of galaxy properties, such as the acquisition of their
angular momentum or their orientation.  
Understanding the way galaxies acquire their spin and shapes can thus
shed light on galaxy formation. 
Intrinsic  galaxy alignments  are  also a  source  of systematics  for
upcoming  lensing  surveys  (see \citealt{2015SSRv..tmp...65J}  for  a
recent review).  
 
\smallskip

The    tidal    torque   theory    \citep[TTT,][]{1951pca..conf..195H,
  1970Afz.....6..581D, 1984ApJ...286...38W,2000ApJ...532L...5L} states 
that protogalaxies acquire their angular momentum by the gravitational
torque due to  the misalignment of their inertia tensor  and the tidal
tensor due to the large-scale structures.  
It then predicts that the spin should be aligned with the intermediate
axis (orthogonal  to the filaments and  in the plane of  the walls) of
the tidal tensor.
However, this theory is only valid  in the linear to mildly non-linear
regime,  and  cannot  predict  accurately the  eventual  spin  of  the
collapsed halo \citep{2002MNRAS.332..325P}.

\smallskip

$N$-body simulations have shown that haloes in walls have their spins in
the plane of the  wall, confirming the TTT \citep{2007MNRAS.381...41H,
  2007ApJ...655L...5A}. 
However,  they  also  showed  that   the  situation  in  filaments  is
mass-dependent.  
Low-mass haloes  have their  spins aligned with  the direction  of the
filaments,      while      massive       ones      are      orthogonal
\citep{2007MNRAS.381...41H,  2007ApJ...655L...5A, 2012MNRAS.427.3320C,
  2012MNRAS.425.2049H, 2013ApJ...766L..15L, 
  2013ApJ...762...72T, 2014MNRAS.440L..46A}.  
In   a    detailed   study    of   the    halo--filament   alignments,
\citet{2012MNRAS.427.3320C} argued that low-mass haloes build their 
mass from smooth accretion along the filaments, and therefore acquire a
spin parallel to it, while the  more massive ones undergo mergers that
tend to build a spin orthogonal to the filament.
\citet{2015MNRAS.446.2744L} showed that the origin  of the spin is due
to the vortices in the filaments.

\smallskip

Simulations showed  that the  major axis  of the  halo has  a stronger
alignment signal with the LSS (direction  of filament, or in the plane
of   the    wall)   \citep{2007ApJ...655L...5A,   2009ApJ...706..747Z,
  2013MNRAS.428.2489L,  2014MNRAS.443.1090F},   the  alignment  signal
being stronger for more massive haloes.  
Using the tidal web, \citet{2014MNRAS.443.1090F} found a 
strong alignment of the major axis of the haloes with the direction of 
the filaments  and the  normal of  the walls as  defined by  the tidal
field. 
Interestingly, when  defining the  LSS using a  mass-weighted velocity
shear tensor, they  found an anti-alignment between the  major axis of
the halo and the direction of the filament or normal to the wall.  

\smallskip

However,  galaxies  are  not   only  sensitive  to  the  gravitational
component, but also governed by baryonic processes.
Thanks to recent progress in modelling the baryonic processes involved
in galaxy formation as well as  an increase of computing power leading
to better resolution, galaxy alignments have also been investigated in
hydrodynamical simulations 
\citep{2015MNRAS.453..721V, 2015MNRAS.454.3328V, 2015MNRAS.454.3341C,
  2015MNRAS.454.2736C, 2015arXiv151200400W,2016MNRAS.460.3772S}. 
Using   the    Horizon-AGN   simulation   \citep{2014MNRAS.444.1453D},
\citet{2015MNRAS.454.2736C} found two alignment signals, corresponding
to elliptical and spiral galaxies. 

\smallskip

Observationally,     the      picture     is     not      so     clear
\citep{2002ApJ...567L.111L,  2005ApJ...629L...5L, 2006MNRAS.369.1293Y,
  2007MNRAS.379.1011A,    2007ApJ...671.1248L,    2009ApJ...703..951W,
  2011ApJ...732...99L, 2013ApJ...779..160Z, 2015ApJ...798...17Z}.  
\citet{2011ApJ...732...99L} found  a stronger correlation of  the spin
of spiral galaxies than predicted  by the TTT, implying a misalignment
between the galaxy and the dark halo.
\citet{2006MNRAS.369.1293Y}  using   the  SDSS  DR2  found   that  the
positions of satellites are aligned with  the major axis of the group,
with a stronger alignment for  red centrals and satellites, for higher
halo mass, and at smaller radii. 
These results  were extended by \citet{2009ApJ...703..951W}  who found
that neighbouring groups tend to be aligned with the major axis of the
target groups.
\citet{2013ApJ...779..160Z}  using the  SDSS  DR7  found an  alignment
signal between the major axis of central galaxies and the direction of
filaments  and the  plane of  walls, while  no signal  was found  for
satellites.  
The alignment  of red  centrals is of  the same order  as that  of the
halo, while blue centrals show a weaker alignment.
\citep{2015A&A...576L...5T} studied the alignment of galaxy pairs with
the filaments, and found that loose pairs ($d>\SI{0.3}{Mpc}$) are more
aligned with the direction of the filaments than close pairs.

\smallskip

However, reliably defining the LSS in observation is a difficult task.
Alternatively,   one   may   look   at  alignments   in   halo   pairs. 
In our previous paper \citep[hereafter Paper I]{2015MNRAS.451..527L}, we
studied  the rate  at  which  targets undergo  an  interaction with  a
neighbour at least 0.4 times as massive.  
In this  study, we used  the same data  and method to  investigate the
alignment of spin and shapes of interacting pairs.

\smallskip

Section~\ref{sec:meth}     briefly     presents     the     simulation
\citep{2015JKAS...48..213K}         and          the         catalogue
\citepalias{2015MNRAS.451..527L}, \S~\ref{sec:res} shows our main
results  in  terms of  alignments  of  spins,  shapes, and  orbits  of
interacting pairs, and the conclusions are drawn in \S~\ref{sec:ccl}.

\section{Simulation and method} 
\label{sec:meth}

\subsection{The Horizon Run 4 simulation}
We used  the Horizon  Run 4 simulation  \citep{2015JKAS...48..213K}, a
massive   $N$-body  simulation   with   $N=6300^3$   particles  in   a
$L=\hGpc{3.15}$  box  in  a  flat  WMAP5  $\Lambda$-cold  dark  matter
cosmology,   starting  at   $z=100$  with   second  order   Lagrangian
perturbation theory, ensuring a 1\%-level accurate  power spectrum and 
halo mass function \citep{2014NewA...30...79L}.
Haloes were  detected using  the Ordinary  Parallel Friends-of-Friends
algorithm \citep[OPFoF,][]{2015JKAS...48..213K}  with a  linking length
of    0.2   times    the   mean    particle   separation,    and   the
gravitationally-bound subhaloes,  which are assumed to  host galaxies,
with the PSB subhalo finder \citep{2006ApJ...639..600K}.
The   PSB    subhalo   finding   method   is    similar   to   SUBFIND
\citep{2001MNRAS.328..726S} using peak finding and density gradient to
allocate members to the  subhalo candidates. 
Additionally,  PSB uses  tidal  boundaries to  demarcate each  subhalo
region.
For more details, we refer the readers to \citet{2006ApJ...639..600K}.
Hereafter, we will refer to PSB subhaloes as haloes.

\smallskip 

\subsection{Catalogue and definitions} 
The catalogue was described in \citetalias{2015MNRAS.451..527L}.  
Our target  (T) catalogue consists  of all haloes more  massive than
$M=\hMsun{5e11}$, while the neighbour  catalogue (N) consists of those
more massive  than \hMsun{2e11}, corresponding respectively  to 56 and
23 particles.  
A target of  mass \Mt\ is defined  to be interacting if  it is located
within the virial radius of its nearest neighbour of mass $\Mn > \xi_0
\Mt$, with $\xi_0 = 0.4$.  
Therefore, the  maximum separation allowed to  an ``interacting'' halo
pair depends on the virial radius of the neighbour halo. 
This choice was made in accordance with the idea that a halo should be
regarded interacting when it is under the significant influence of its
nearest neighbour.
It is also  based on the observational finding that  the effect of the
morphological  type of  the  neighbour starts  affecting  that of  the
target only  when the pair  separation between the members  is shorter
than $R_\text{vir,N}$ \citep[see Fig. 6 of ][]{2009ApJ...691.1828P}.
Note  that in  this  definition, we  are  counting interactions  per
target halo rather than per pair.
Therefore, some interactions  are counted twice, e.g.,  if the target
is also the neighbour's nearest neighbour, but it is not necessarily
the case.

\smallskip

To quantify  the environment, in addition  to the target mass,  we use
the large-scale density smoothed over the 20 nearest neighbours 
\begin{equation}
  \rho_{20} = \sum_{1}^{20} M_i W(r_i/h),
\end{equation}
where $M_i$  is the mass  of the  $i$th neighbour, $r_i$  the distance
between the  target and the  $i$th neighbour, $h$ the  smoothing scale
chosen  to  enclose the  20  closest  neighbours,  and $W$  the  cubic
B-spline  smoothing  kernel  used  in  smooth  particle  hydrodynamics
simulations \citep{1985A&A...149..135M}.  
The density is  then normalized to a  dimensionless parameter $\delta$
by
\begin{equation}
  1+\delta = \frac {\rho_{20}} {\bar\rho},
\end{equation}
where 
\begin{equation}
  \bar \rho = \frac 1 V \sum_{i\in\text{N}}M_i
\end{equation}
is the mean density of the neighbour catalogue.
The choice of 20 neighbours has proven to keep the noise in the smooth
density  small while  allowing one  to reach  a small  smoothing scale
(\citealt{2007ApJ...658..898P,                   2008ApJ...674..784P};
\citetalias{2015MNRAS.451..527L}).  

As in  \citetalias{2015MNRAS.451..527L}, we define three  mass bins at
each redshift (two  at $z=3.1$ and one at $z=4$)  with the same number
of targets, and subdivide them into 3 density bins so that each of the
9 (respectively 6 or 3) bins has the same number of targets.  
This  introduces  three  mass   and  six  density  thresholds  $M_i(z),
\Delta_{j,i}(z);\, i\in  \{0,1,2\}, j\in\{1,2\}$, that can  be seen in
Figure~A1 of \citetalias{2015MNRAS.451..527L}.
The values  of $M_0,M_1,M_2$ at $z=0$  are respectively \num{7.68e12},
\num{1.10e13}, and \hMsun{1.98e13}.  

Therefore, we stress here that the  results obtained in this paper are
based on a  constant number density of the halo  catalogue rather than
fixed mass bins.
In \S~\ref{sec:align_md}, we address the issue of the mass and density
dependence using fixed bins.

\subsection{Characterising the alignment signal}
\label{sec:count}
In order  to detect an  alignment signal for  a given angle  $\theta =
(\vect u, \vect v)$ between any  two vectors $u$ and $v$ associated to
a   halo,    we   used   the   normalized    pair   counts   following
\citet{2005ApJ...628L.101B} and \citet{2006MNRAS.369.1293Y}, comparing
the measured number  of pairs with the expected count  from the random
case.  
\begin{itemize}
  \item We counted $N(\theta)$ the number of pairs for a given $\theta$
  \item We then randomly reorder \vect u $100$ times
  \item    We   calculated    the   mean    and   standard    deviation
    $\mean{N^\text{R}(\theta)}$ and $\sigma_\theta$
  \item  We considered  the  normalized pair  count  $f(\theta) =  \tfrac
    {N(\theta)}{\mean{N^\text{R}(\theta)}}$ 
  \item  The  strength  of  the   signal  (error  bars)  is  given  by
    $\tfrac{\sigma_\theta }{\mean{N^\text{R}(\theta)}}$ 
\end{itemize}

In  three   dimensions,  a  uniform  distribution   of  angles  yields
$f(\cos\theta) =  1$, therefore  we used bins  with constant  width in
$\cos\theta$.   A  value  of  $f(\cos\theta\simeq 0)\gg  1$  shows  an
anti-alignment (orthogonality),  while $f(\cos\theta  = \pm 1)  \gg 1$
shows an alignment.  
  
\section{Results} 
\label{sec:res}

\subsection{Distribution of spins and shape parameters}

\subsubsection{Spin parameter}

\begin{figure}
  \centering
  \includegraphics[width=\columnwidth]{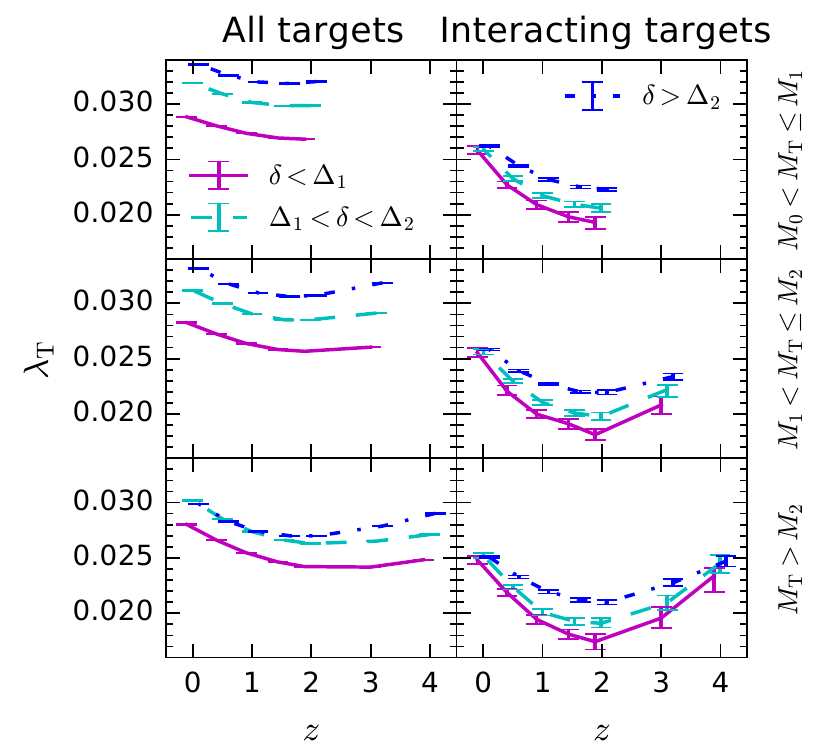}
  \caption{\label{fig:spin}%
    Spin parameter of all (left column) and interacting (right column)
    targets  for  target   masses  $M_0<\Mt<M1$  (top),  $M_1<\Mt<M_2$
    (middle),  and  $\Mt>M_2$  (bottom panel),  and  $\delta>\Delta_2$
    (dash-dotted  lines),  $\Delta_1<\delta<\Delta_2$ (dashed  lines),
    and $\delta>\Delta_2$ (solid line).  
    The  lines are  median values,  and the  error-bars show  the 95\%
    confidence intervals obtained by bootstrapping 1000 times.  
  }
\end{figure}

The rotation  of haloes  can be  quantified by  the spin  parameter as
defined by \citet{1969ApJ...155..393P}: 
\begin{equation}
  \label{eq:spin}
  \lambda = \frac{|\vect{J}|\sqrt{|E|}}{GM^{5/2}},
\end{equation}
where $\vect J$ is the sum of  the angular momenta of each particle in
the  halo, and  $E =  W +  K$ is  the total  (kinetic plus  potential)
energy\footnote{We  note that  the  potential energy  $W$ is  computed
  using  the smooth  potential as  calculated in  the simulation.  The
  calculation  of  the  potential   may  affect  the  calculated  spin
  parameter \citep{2014JKAS...47...77A}.}.  
This corresponds to the ratio of the rotation to the random motions.  
Galaxies with  purely random motion  have a  spin parameter of  0, and
close to unity for fully rotationally-supported galaxies.

We  studied  the  distribution  of the  spin  parameter  $\lambda$  of
interacting  targets in  the 9  bins of  mass and  density defined  in
\S~\ref{sec:meth}.  
Figure \ref{fig:spin} shows the median of the distribution of the spin
parameter of interacting haloes for  the three mass bins $M_0<\Mt<M_1$
(top), $M_1<\Mt<M_2$  (middle), and $\Mt>M_2$ (bottom  panel), for all
(left-hand column) and interacting (right-hand column) targets.  
In each  panel, we  show the  median of each  subsamples in  the three
density   bins   $\delta>\Delta_2$,  $\Delta_1<\delta<\Delta_2$,   and
$\delta<\Delta_1$.  
The  error bars  on  the  median show  the  95\% confidence  intervals
obtained by bootstrapping 1000 times.  

\begin{figure}
  \centering
  \includegraphics[width=\columnwidth]{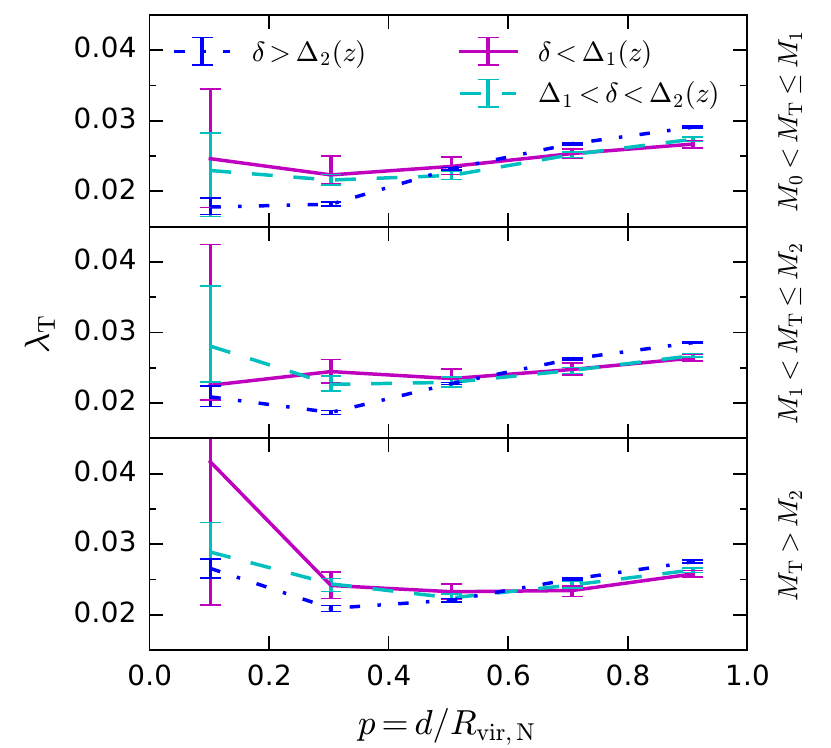}
  \caption{\label{fig:spin_p}%
    Spin parameter of interacting targets (at $z=0$) as a function of the
    normalized separation $p=d/R_\text{vir}$. 
  }
\end{figure}

In  the left-hand  panels, the  spin parameters  of all  targets first
decrease until $z\simeq 2$, then increase until $z=0$.
At fixed  mass, targets in  high-density regions have a  higher median
spin than the lower-density ones.  
However, we  note that targets  in the top  panel have fewer  than 300
particles at $z>1.5$; while targets in  the middle and bottom panel have
fewer than 300 particles for $z>2$.
Therefore, the  discreteness effect may  affect the estimation  of the
spin  \citep{2007MNRAS.376..215B}, 
{%
  therefore the evolutions at $z>1.5$ (top panel) or 2 (middle and
  bottom panels) may not be reliable.
}
For haloes more  resolved than 300 particles, the median  value of the
spin is increasing with decreasing redshift.
The typical increase  between $z=2$ and 0 is 7\%  in the low-mass bin,
and up to 15\% in the high-mass and low-density bin.

\begin{figure*}
  \centering
  \begin{subfigure}[b]{.48\textwidth}
    \includegraphics[width=\columnwidth]{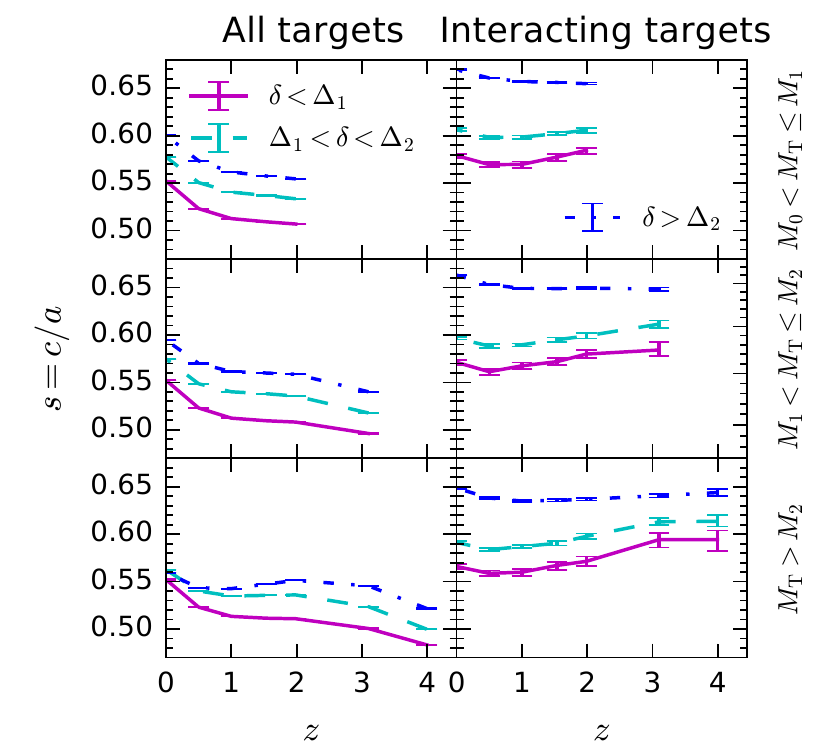}
    \caption{\label{fig:sshape_all}%
      Sphericity $s=c/a$
    }
  \end{subfigure}  
  \begin{subfigure}[b]{.48\textwidth}
    \includegraphics[width=\columnwidth]{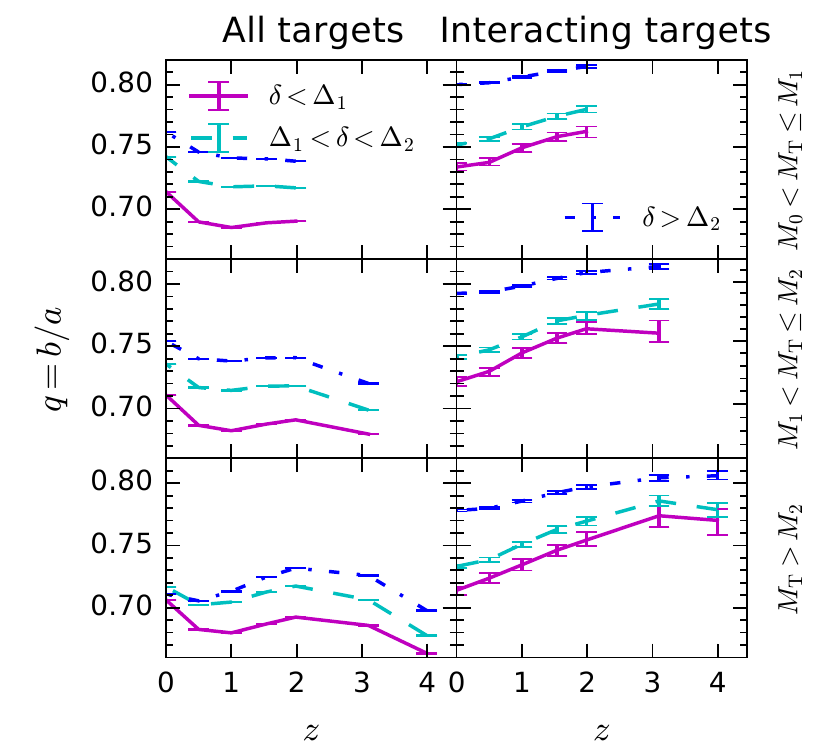}
    \caption{\label{fig:qshape_all}%
      Oblateness $q=b/a$
    }
  \end{subfigure}  
  \caption{\label{fig:shape_all}%
    Sphericity (left)  and oblateness (right) of  interacting targets;
    same legend as Fig.~\ref{fig:spin}. 
  }
\end{figure*}

The  very weak  evolution of  the spin  with mass  is consistent  with
previous  studies,  that did  not  find  any  dependence of  the  spin
parameter distribution on mass \citep{2007MNRAS.376..215B}.  
However, the very  large volume and good statistics of  HR4 enabled us
to  detect  a  weak  but   significant  signal  of  the  influence  of
interactions in the distribution of the spin parameter.
In the right-hand  panels, the spin parameter  of interacting targets
shows a similar behaviour to that of all targets:
It also  reaches a minimum  at $z\simeq  2$, and increase  until $z=0$
where it  interestingly has  a similar  value at all  bin of  mass and
density, $\lambda_\text{T}\simeq  0.025$.
Here  again, we  cannot conclude  about  the upturn  since haloes  are
poorly resolved at $z>2$.
At $z=2$, in the higher mass bin, the
median  varies from  0.0175 in  the  low-density regions  to 0.021  in
high-density.  
The typical  increase of $\lambda_\text{T}$ of  interacting targets is
larger than that of all targets. 
Between  $z=2$ and  0,  it increases  by 17  to  19\% in  high-density
regions,  26 to  32\%  in  intermediate regions,  and  33  to 42\%  in
low-density regions.

The spin parameter  of interacting target is smaller than  that of all
targets.
This can be understood by the tidal forces from the neighbour they are
interacting with.
This  can be  seen in  Figure~\ref{fig:spin_p}, which  shows the  spin
parameter of  interacting targets at redshift  0 as a function  of the
separation  normalized by  the virial  radius  of the  neighbour $p  =
d/R_\text{vir}$.  
The spin  parameter decreases  with decreasing separation,  from about
0.025 at $p=1$ to about 0.02  at $p\simeq 0.25$, which supports the idea
that  interactions tend  to  slow  down the  targets,  since the  spin
parameter is smaller at smaller separation, where the tidal forces are
stronger.  
At  very-low $p<0.3$,  there  seems  to be  an  upturn  in the  spin
parameter.
This effect is  not significant, given the  large error-bars, except
in the higher mass bin, but it seems to be systematic. 
Interestingly,  \citet{2012MNRAS.426.1606C},   who  used   the  same
definition  of  neighbours  as  ours,  found  similar  results  from
observations while measuring the spin  of spiral galaxies which have
an early-type neighbour.

  According  to  the TTT,  the  two-point  spin correlations  show  an
  increasing correlation  with a smaller  separation (see eq.   16 and
  Figure 11 in \citet{2002MNRAS.332..325P}).
  However, this  trend is lessened  at lower redshifts, mainly  due to
  the effects of nonlinear evolution.
  This may indicate that the positive proximity effect on the target's
  spin amplitude comes from the tidal interactions with the neighbour
  halo, which is stronger as the separation becomes smaller.

To our knowledge, this is the first numerical study of the effects of
interactions on the spin of dark matter haloes.

\subsubsection{Shape parameters}

\begin{figure*}
  \centering
  \begin{subfigure}[b]{.48\textwidth}
    \includegraphics[width=\columnwidth]{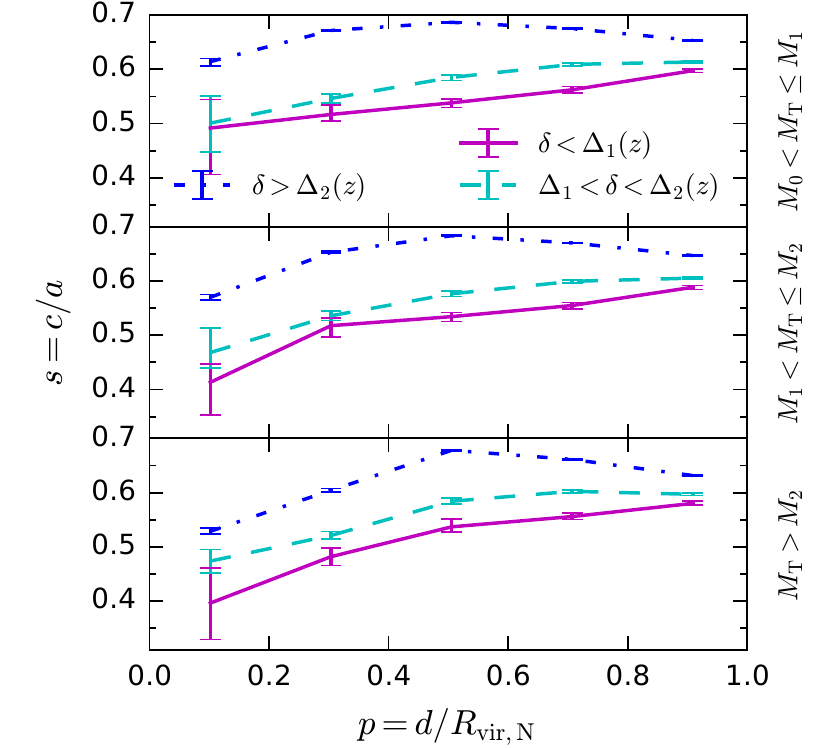}
    \caption{\label{fig:sshape_p}%
      Sphericity $s=c/a$
    }
  \end{subfigure}  
  \begin{subfigure}[b]{.48\textwidth}
    \includegraphics[width=\columnwidth]{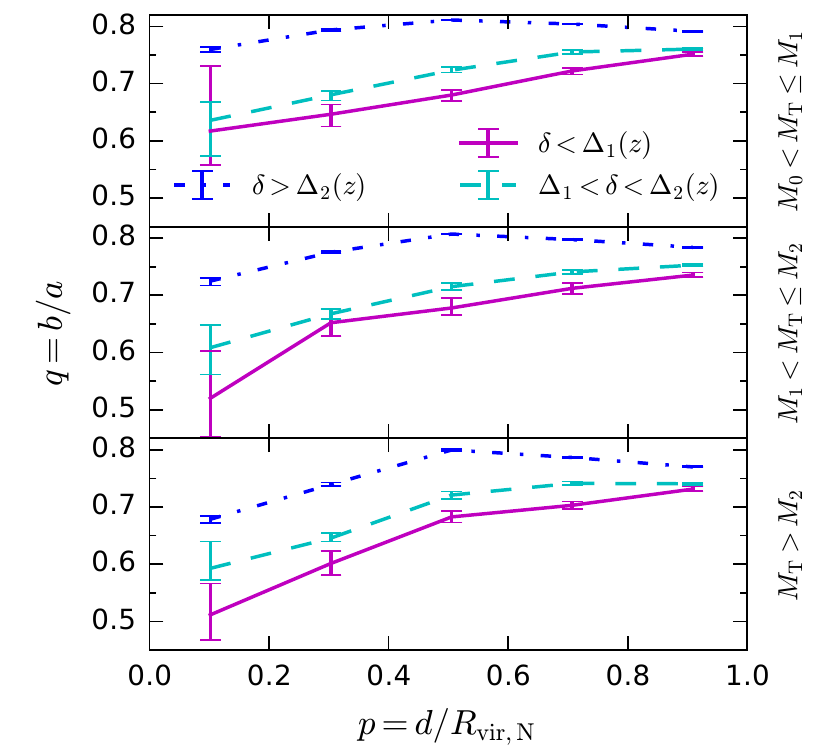}
    \caption{\label{fig:qshape_p}%
      Oblateness $q=b/a$
    }
  \end{subfigure}  
  \caption{\label{fig:shape_p}%
    Sphericity and oblateness of interacting  targets as a function of
    the normalized separation $p=d/R_\text{vir}$ at $z=0$;
    same legend as Fig.~\ref{fig:spin}.
  }
\end{figure*}

Haloes are usually not spherically symmetric. 
To quantify  their triaxiality, we computed  the pseudo-inertia tensor
\tens I defined as 
\begin{equation}
  \label{eq:shape}
  I_{ij} = \sum_\alpha x_{\alpha,i} x_{\alpha,j},
\end{equation}
where $x_{\alpha,i}$ is  the $i$th component of  the relative position
of particle $\alpha$ with respect to the centre of mass of the halo.  
This tensor has three positive eigenvalues $a^2\geq b^2 \geq c^2$, and
the  corresponding three  eigenvectors $\vect  a, \vect  b, \vect  c$,
respectively referred to as the major, intermediate, and minor axes.  
{%
  The choice of this tensor, rather than the reduced inertia tensor 
  (where the contribution  of each particle is  weighted by $1/r^2$)
  is  motivated  by  the  fact  that   it  is  easier  to  compare  to
  observation,   and   more   relevant   to   weak   lensing   studies
  \citep{2012MNRAS.420.3303B}. 
  Moreover, since PSB  subhaloes do not include the mass  of their own
  subhaloes, the inertia is not contaminated by the latter.

  The effects of  the weighting scheme on the results  is discussed in
  \S~\ref{sec:weight}. 
}

The sphericity $s$, oblateness $q$, and prolateness $p$ parameters are
defined as 
\begin{equation}
  \label{eq:qs}
 s =  {\frac c a}\mbox{,  }q = {\frac  b a}\mbox{, and  }p =
 {\frac c b}. 
\end{equation}

Haloes are said to be respectively spherical, oblate, or prolate, if
\begin{subequations}
  \begin{align}
    a\simeq b\simeq c,&&& \mbox{or } s\simeq 1,    \label{eq:shapes}\\
    a \simeq  b \gg c,  &&& \mbox{or  } q \simeq  1\mbox { and  } s\ll
    q, \label{eq:shapeq}\\ 
    a\gg b\simeq  c, &&&  \mbox{or }  p \simeq  1 \mbox  { and  } q\ll
    1, \label{eq:shapep} 
  \end{align}
\end{subequations}
and triaxial in other cases.

Figure          \ref{fig:shape_all}\subref{fig:sshape_all}         and
\subref{fig:qshape_all}  respectively show  the  median  value of  the
sphericity and  oblateness parameters  of all (left-hand  columns) and
interacting (right-hand  columns) targets  as a function  of redshift,
for the same bins of mass and density as Fig.~\ref{fig:spin}.  
In the all-target case (left-hand column), the sphericity increases  with
decreasing redshift at all  bin of mass
and density, from  about 0.5 to 0.55: haloes become  more spherical at
lower redshifts.
The oblateness increases from $z=4$  to 2, then decreases until $z=1$,
and increases again until $z=0$.
At fixed  redshifts and mass,  for the all-targets case, $q$  and $s$
increase with density.
Finally, at fixed density and  redshift, the sphericity is independent
of mass in the mass range probed by our constant number density bins.

These results are in agreement with previous results, for instance,
\citet{2012JCAP...05..030S} who  studied the distributions of  $s$ and
$q$ in the Millennium I and II simulations \citep{2005Natur.435..629S,
  2009MNRAS.398.1150B}. 
However, the  very good  statistics of  the HR4 enables  us to  see an
additional dependence on the large-scale density.

The                right-hand                 panels                of
Figs.~\ref{fig:shape_all}\subref{fig:sshape_all}                   and
\subref{fig:qshape_all}  show  the  evolution of  the  sphericity  and
oblateness  parameters   of  those  target  that   are  undergoing  an
interaction.  
The sphericity and  oblateness of interacting targets  are higher than
all  targets:  both  are  significantly  higher  for  the  interacting
targets. 
Moreover, their  redshift-evolution is  different: both  decrease with
decreasing redshift, as opposed to the all-target case.

The  dependence   of  $s$  and   $q$  on  the   normalized  separation
$p=d/R_\text{vir}$       at       $z=0$        is       shown       on
Fig.~\ref{fig:shape_p}\subref{fig:sshape_p} and~\subref{fig:qshape_p}.  
In both cases, the parameter decreases with decreasing separation.
However, it  is interesting to  notice that in high-  and intermediate
densities, the  parameter reaches a maximum,  respectively at $p\simeq
0.5$ and 0.7, while at low-density, the decrease is monotonous. 
In the latter  case, the maximum is presumably located  outside of the
virial radius of the neighbour.

{%
This is surprising at first, but can be understood by these two facts:
(1) the number of interactions decreases with $p$, so we are dominated
by large-$p$ interactions  (loose-pairs), which tend to  have a larger
sphericity and oblateness.  
(2) In  the high-density bin, the  maximum $s$ and $q$  are reached at
$p\simeq 0.5$, then both quantities decrease.
In the  intermediate-density bin, the  maximum seems to be  reached at
$p\simeq 0.9$.  
It is reasonable to believe that for the lower bin of density, similar
behaviour can be seen beyond \rvir. 
Therefore,  non-interacting targets,  which are  located at  more than
$1\rvir$ from the nearest neighbour, must have a lower $s$ and $q$. 
}

Halo  shapes ($c/a$  and $b/a$)  as shown  in Figure~\ref{fig:shape_p}
become  more elongated  for  closer  pairs, which  may  come from  
stronger tidal  interactions (their  shapes are stretched  toward each
other).
This  trend  becomes  less  obvious   in  denser  regions,  where  the
interaction  with the  closest neighbour  becomes less  important with
respect to interactions with other neighbouring halos.
The  effects  from  other  halos  may  thus  distort  the  pair  tidal
interactions.

\begin{figure}
  \begin{center}
    \begin{tikzpicture}
      \filldraw[color=blue!60, fill = blue!5,thick]
      (0,0)ellipse(1.5 and 0.5); 
      \filldraw[color=red!60,    fill   =    red!5,   thick,    rotate
      around={30:(2,2)}] %
      (2,2) ellipse (1.5 and 0.5); 
      \draw [->, color=blue!50,thick] (0,0)--(2,0) ; 
      \draw [->, color=blue!50,thick] (0,0)--(2,2) ; 
      \draw [dashed, color=blue!50,thick] (2,2)--(3,3) ; 
      \draw [->, color=red!50,thick] (2,2)--(3.73205080757,3) ; 
      \node[anchor=east,color=blue] (T) at (0,0) {T};
      \node[anchor=east,color=red] (N) at (2,2) {N};
      \draw[->, color = blue!50,thick](1,0) arc (0:45:1);
      \node[color=blue] (gamma) at (.8,.2) {$\gamma$}; 
      \draw[->, color = red!50,thick](2.7071,2.7071) arc (45:30:1);
      \node[color=red] (epsilon) at (2.9,2.7) {$\eps$}; 
      \node[anchor=south,color=blue] (at) at (2,0) {$\mathbf{a}_\text{T}$}; 
      \node[anchor=south, color=red] (an) at (3.73205080757,3)
      {$\mathbf{a}_\text{N}$}; 
      \node[anchor=north,color=blue] (at) at (2,2) {$\mathbf{r}$}; 
      \node[anchor=south,color=blue] (ct) at (0,1) 
      {$\vect c_\text{T}$}; 
      \draw[->, color = blue!50,thick] (0,0)->(0,1);
      \node[anchor=south,color=blue] (ct) at (-1,1.732) 
      {$\vect J_\text{T}$}; 
      \draw[->, color = blue!50,thick] (0,0)->(-1,1.732);
      \draw[->,color=blue!50,thick] (0,.75) arc (90:120:.75);
      \node[color=blue] (psi) at (-.3,1) {$\psi$};
    \end{tikzpicture} 
    \caption{\label{fig:def_gpsi}Definition  of the  angles $\gamma  =
      (\vect{a}_\text{T},\vect{r})$, $\eps = (\vect {a}_\text{N},\vect
      r)$, and $\psi = (\vect{c}_\text{T},\vect{J}_\text{T})$.  
    }
  \end{center}
\end{figure}
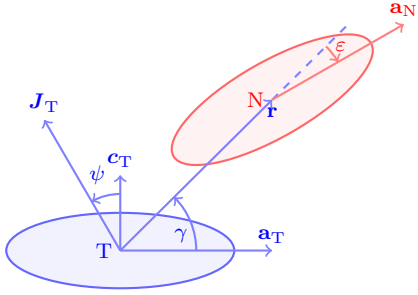

\begin{figure}
  \begin{center}
    \includegraphics[width=\columnwidth]{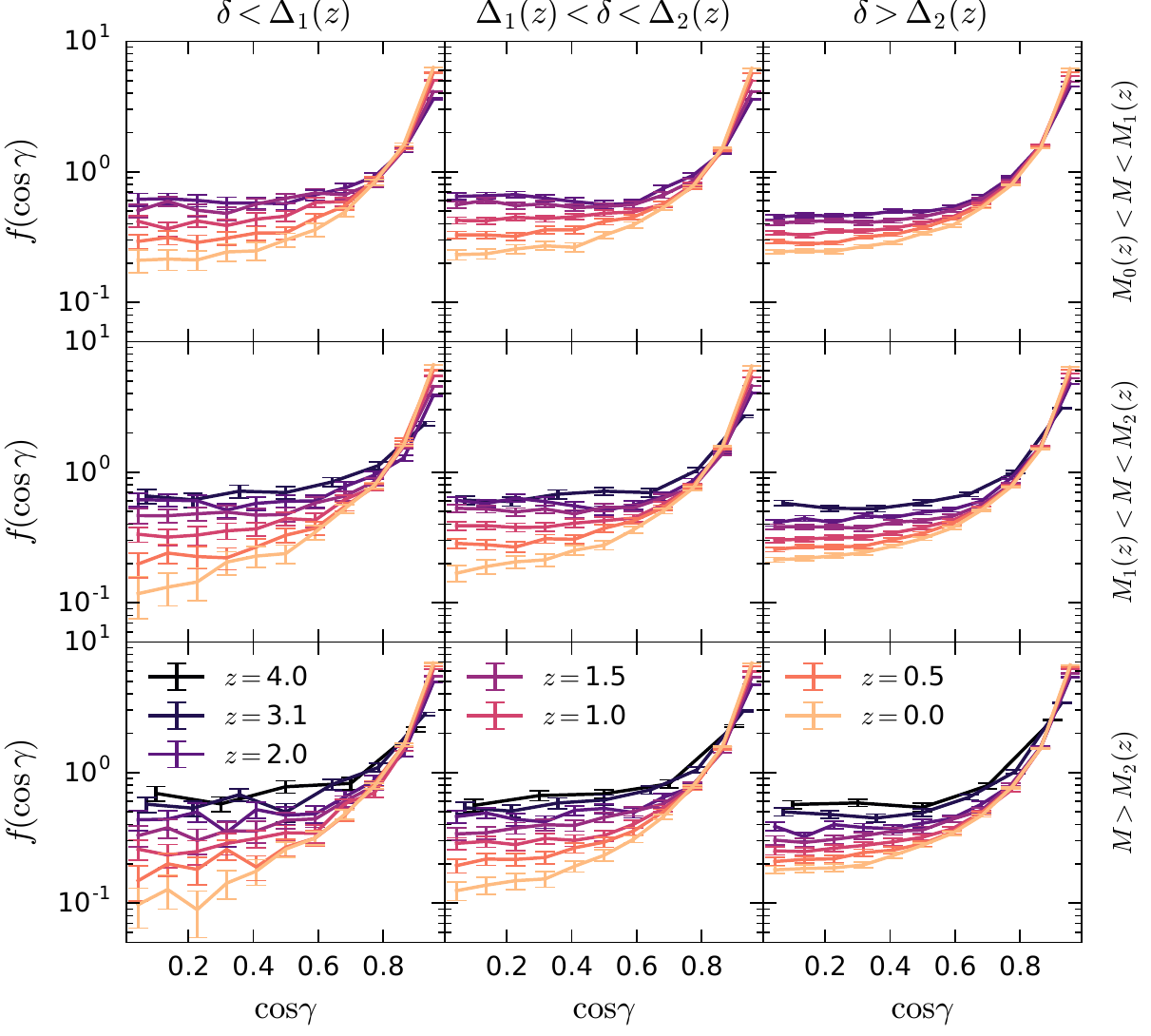}
    \caption{\label{fig:cosgamma}
      Time  evolution  of  the  distribution  of  $\cos\gamma$,  where
      $\gamma = (\vect a_\text{T},\vect{r})$ is the direction of
      the nearest neighbour halo relative to the major axis of the target.
      The  three  rows  correspond  to  the  three  mass  bins  as  in
      \ref{fig:spin}, while the three  columns correspond to the three
      density bins.
      The   error-bars   are   obtained    by   shuffling   the   pairs
      (c.f. \S~\ref{sec:count}). 
    }  
  \end{center}
\end{figure}

\begin{figure}
  \begin{center}
    \includegraphics[width=\columnwidth]{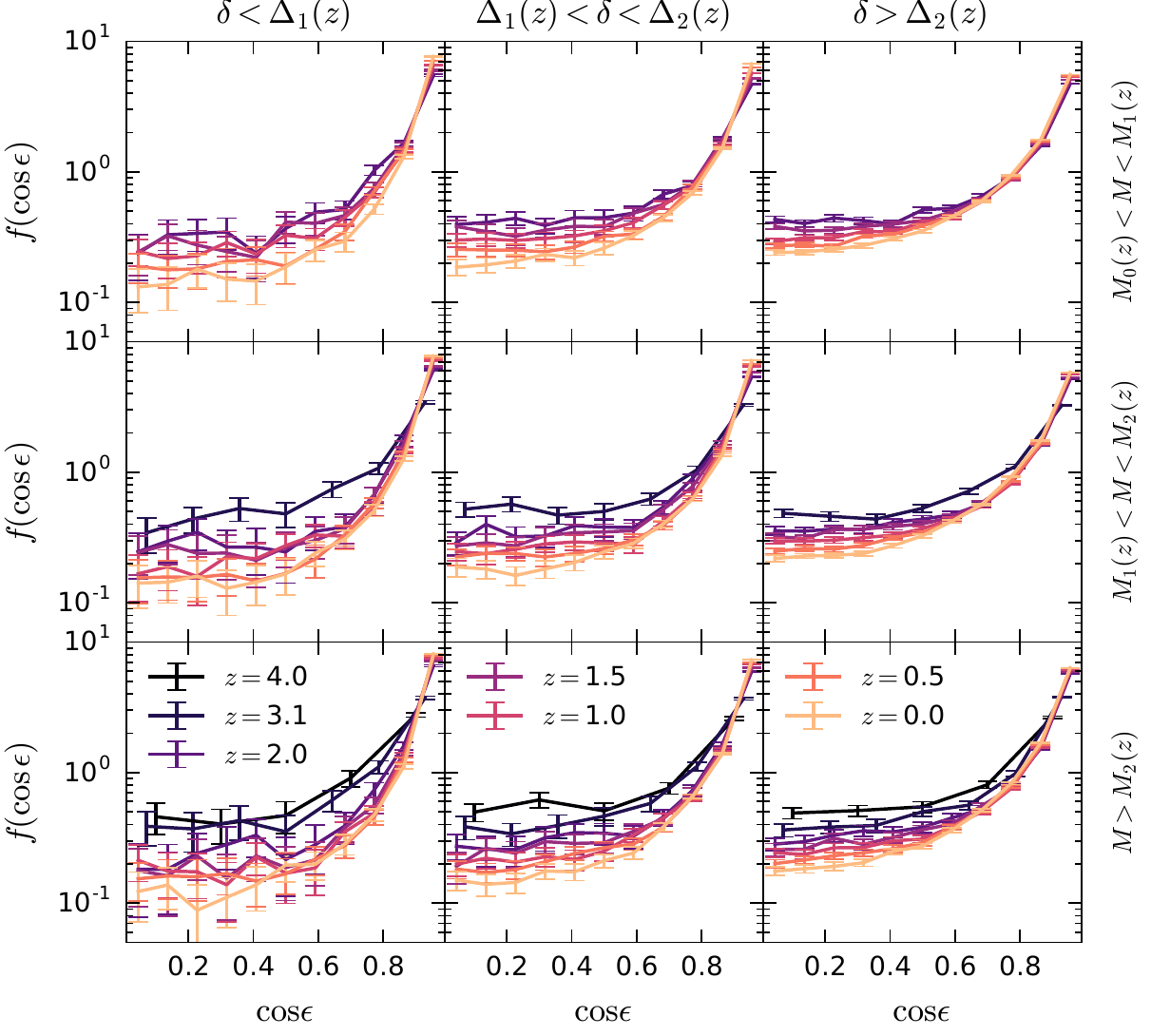}
    \caption{%
      \label{fig:coseps}%
      Time evolution of the distribution  of $\cos\eps$, where $\eps =
      (\vect{  a  _ \mathrm  {N}},  \vect{  r  }  )$ is the angle between
      the halo  separation vector  and the major  axis of  the nearest
      neighbour halo. 
      The legends are the same    as  
      Fig. \ref{fig:cosgamma}.  
    }  
  \end{center}
\end{figure}

\subsection{Alignment of the principal axes}

We studied the alignment of the major axis $\vect a$ associated with
the largest eigenvalue $a$ of the  target and neighbour.
In the following,  we study the angles $\gamma  = (\vect {a}_\text{T},
\vect r)$ and  $\eps = (\vect a_\text{N}, \vect r)$,  where \vect r is
the position vector  of the neighbour with respect to  the target (see
Fig.~\ref{fig:def_gpsi}).  
Note  that these  two angles  are  defined modulo  $\pi/2$, since  the
direction  of the  major axis  is irrelevant,  therefore we  study the
distribution of $\abs{\cos\gamma}$ and \abs{\cos\eps} between 0 and 1.  
Moreover,  in order  to remove  noise from  poorly-defined shapes,  we
require the oblateness  parameter to be smaller than 0.8,  so that the
major axis is well-defined.
This effectively excludes spherical and oblate haloes.
Again, in the lower mass bin, haloes have more than 300 particles from
$z=2$, and from $z=1.5$ in the intermediate and higher mass bins.

Fig.~\ref{fig:cosgamma}  and~\ref{fig:coseps}  respectively  show  the
normalized  pair   counts,  as  defined  in   \S~\ref{sec:count},  for
\abs{\cos\gamma}  and \abs{\cos\eps}  in  the nine  bins  of mass  and
density previously defined.
The  error-bars   were  obtain  following  the   method  described  in
\S~\ref{sec:count}. 
Both figures show very similar features.
Note however that, even though  the definitions of the angles $\gamma$
and  $\eps$  are  symmetric,  the targets  and  neighbours  belong  to
different  catalogues,  so  we  do not  necessarily  expect  the  same
behaviour.
Namely, for a given target T, the  neighbour of its neighbour N is not
necessarily T. 
There  is a  clear  alignment  signal at  all  redshifts, masses,  and
densities, meaning that the neighbour is preferentially located in the
direction of the major  axis of the target, and that  the main axis of
the neighbour is well aligned with the direction to the target. 
At fixed  mass, the  signal is more  significant at  higher densities,
since the statistics are better.  
However, the alignment itself seems to be stronger at low densities.  
Given the small range of mass  probed here, no dependency can be
seen.
The mass  and density dependency will  be studied in more  detail in
\S~\ref{sec:align_md}.  
The signal is weaker at higher redshifts, and becomes stronger as time
passes. 
In addition, the alignment signal  evolves more in low-density regions
than in high density.

\begin{figure}
  \begin{center}
    \includegraphics[width=\columnwidth]{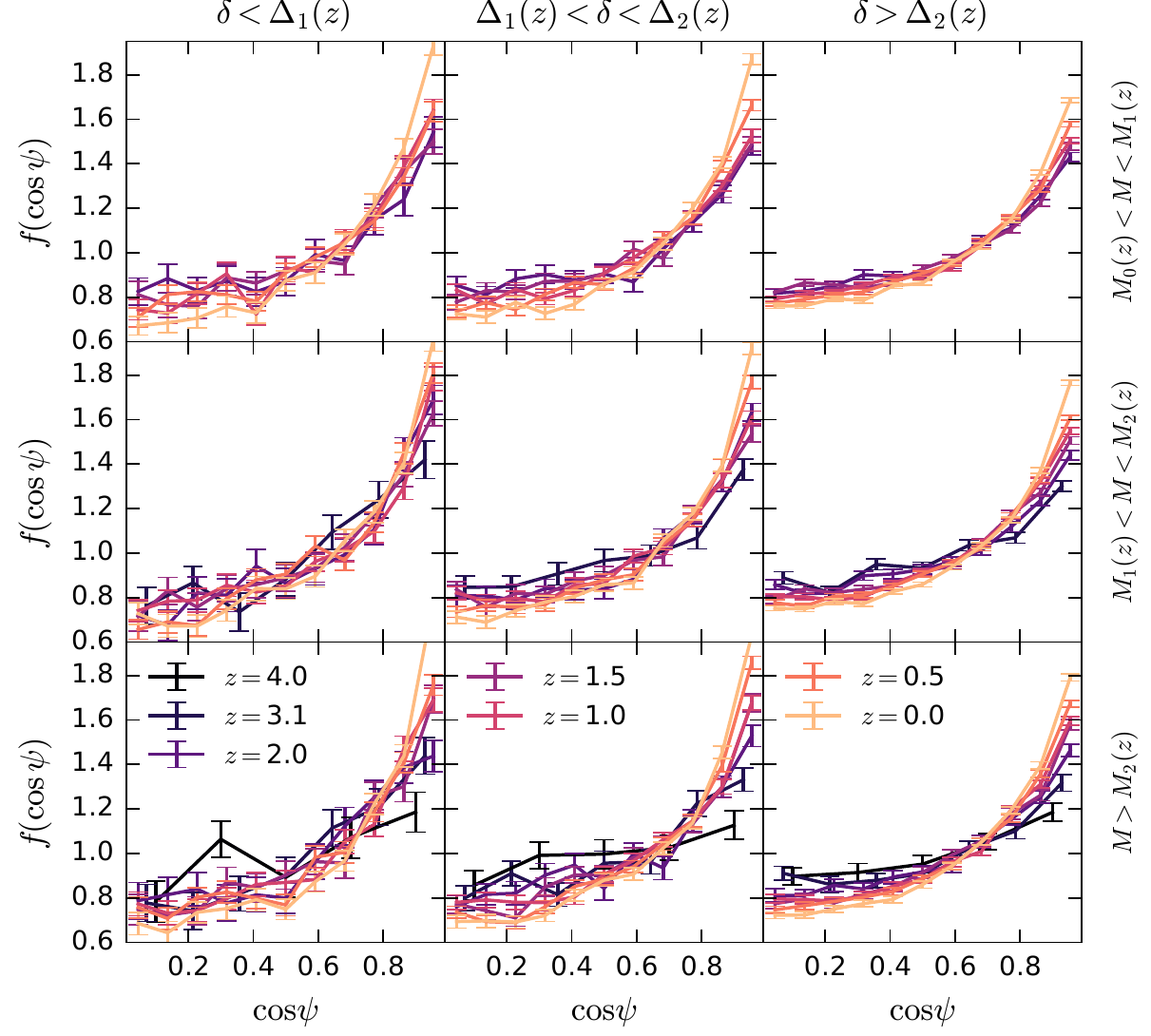}
    \caption{%
      \label{fig:cospsi}%
      Time evolution of the distribution  of $\cos\psi$, where $\psi =
      (\vect{ c _ \mathrm {T}}, \vect{ J_ \mathrm{T} } )$ is the angle between
      the minor axis of the target halo and its angular momentum.
      The  legends are the same        as Fig. \ref{fig:cosgamma}.  
    }  
  \end{center}
\end{figure}

We also studied the alignment between the minor axis of the target and
its spin,  as defined by  the angle  $\psi = (\vect  c_\text{T}, \vect
J_\text{T})$.  
Similarly to  the major axis case,  in order to ensure  that the minor
axis is  well-defined, we require  the prolateness to be  smaller than
0.8. 
Figure \ref{fig:cospsi} shows the normalized pair count for $\psi$ for
the previously-defined mass and density bins.  
The minor axis  and spin are well aligned, as  was previously found by
\citet{2007ApJ...655L...5A}.  
The alignment signal increases with  mass, and decreasing redshift and
density.  
Again, it  is interesting to  see that it  is present at  redshifts as
high as 4.

To summarise,  the position and  the major  axis of the  neighbour are
aligned  with  the   major  axis  of  the   target,  constraining  the
interaction within a small solid angle  with respect to the major axis
of the target. 
These  results are  in  good qualitative  agreement with  observations
\citep{2006MNRAS.369.1293Y,2008MNRAS.385.1511W},  who  found that  the
distributions  of satellites  is aligned  with the  major axis  of the
central galaxy.

In order  to illustrate  how prolate and  triaxial pairs  interact, we
proceeded as follows.  
We first considered pairs such  as both $q_\text{N}$ and $q_\text{T} <
0.8$.  
This ensures that the major axis of both members are well-defined. 
This  represents   \num{4212799}  (\num{2383237})  pairs,   or  29.5\%
(28.1\%) of interacting pairs at redshift 0 (1).

\begin{figure*}
  \includegraphics[width=\textwidth]{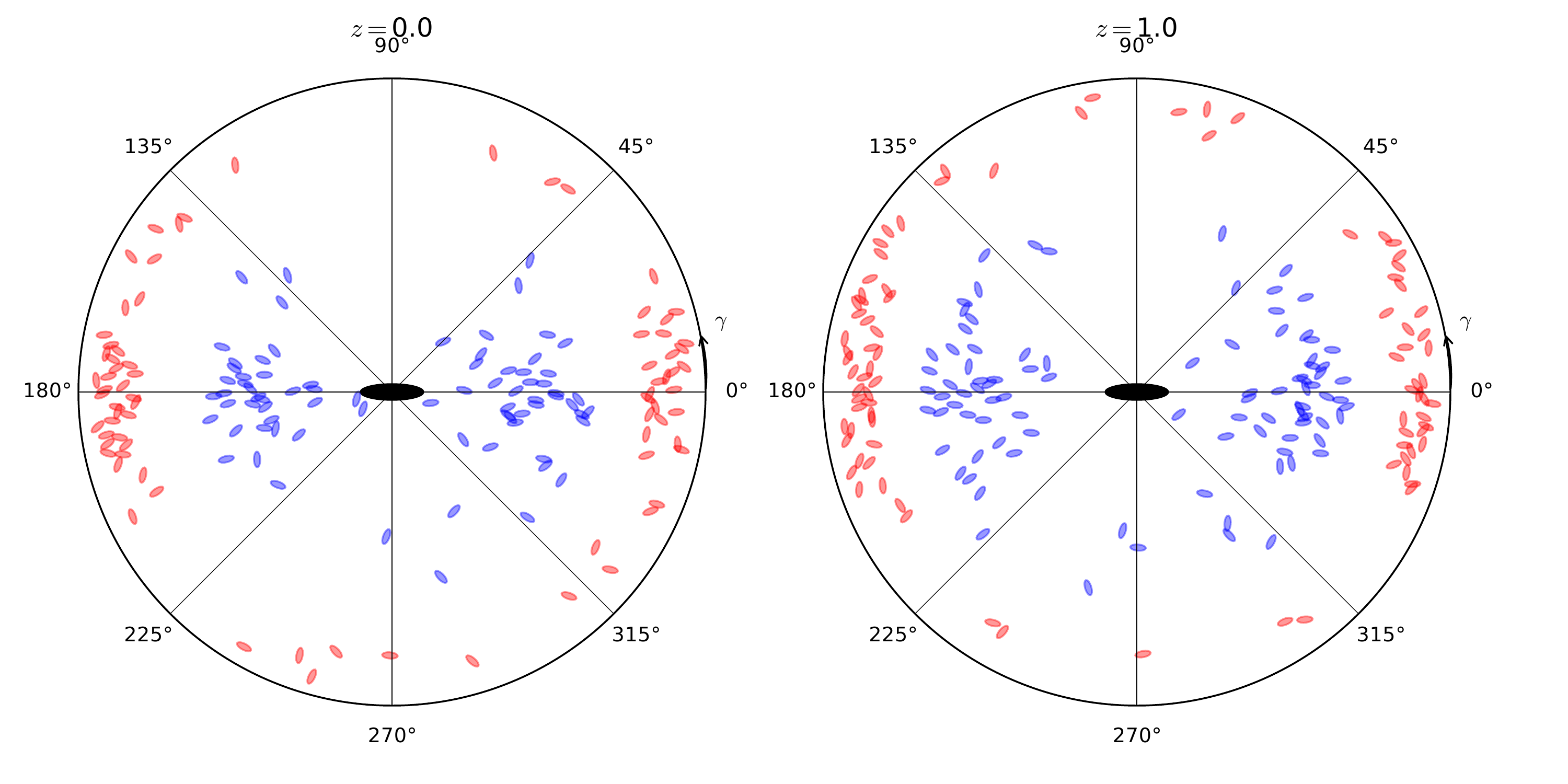}
  \caption{\label{fig:sketch_pro}%
    Distribution of the angular location and orientation of  the nearest neighbour
    with respect to the target.
    The black ellipse at the centre represents the target. 
    Neighbours  are  placed at  positions  $\gamma$  according to  the
    probability distribution of $\gamma$. 
    Note that the neighbours are not necessarily smaller than the target halo.
    The blue and  red ellipses  represent the  neighbours with  their
    orientation with respect to the line connecting it to the target.
    The blue (red)  ellipses show the 33\% closest pairs ($0<p<0.66$), and
    the red ones the 33\% further apart pairs ($0.85<p<1$).  
    For a given $\gamma$, the  orientation of the neighbour $\eps$ and
    the  normalized  separation  $p$  are chosen  randomly  among  all
    neighbours at this $\gamma$. 
  }
\end{figure*}

\begin{figure}
  \includegraphics[width=\columnwidth]{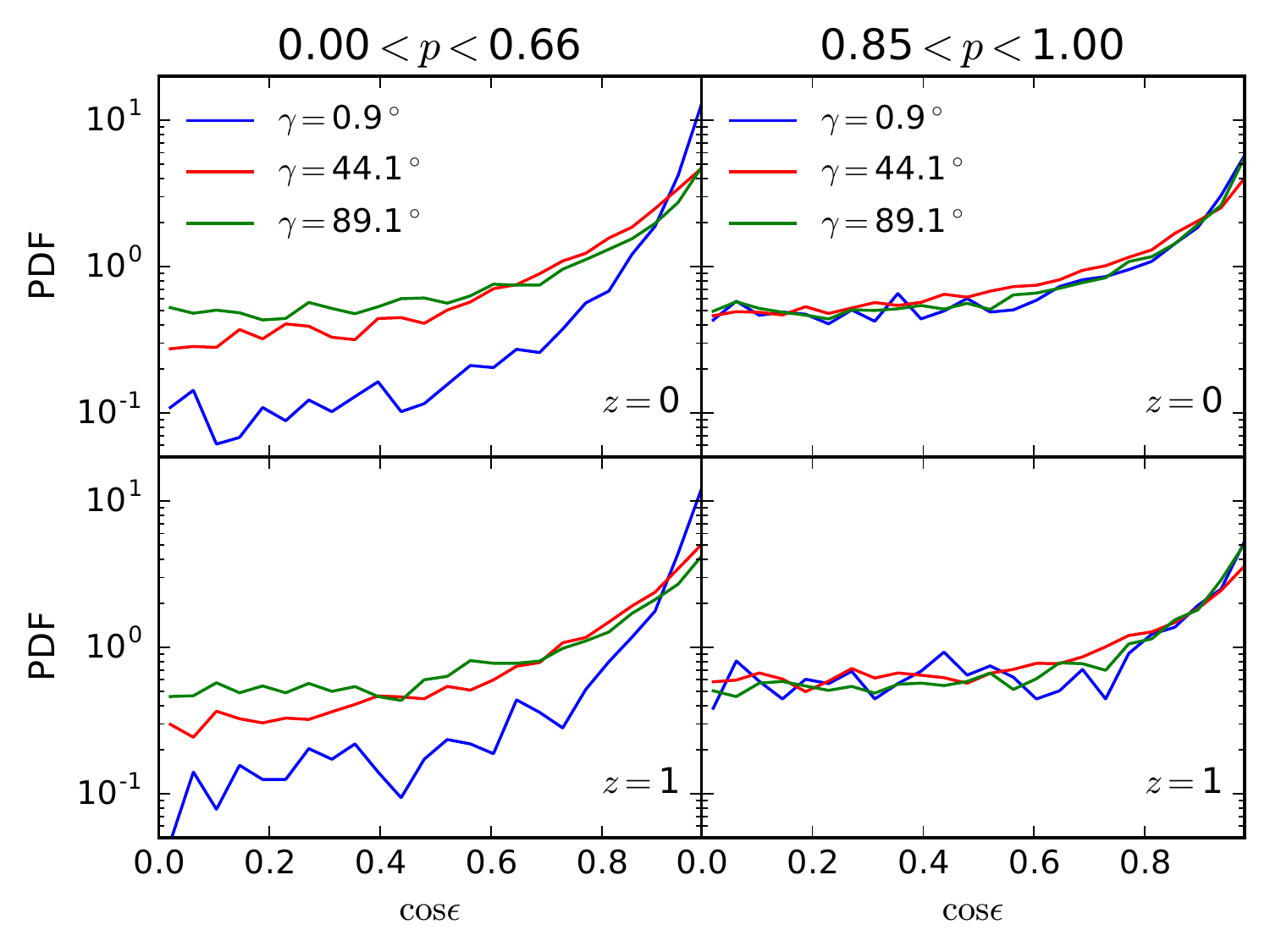}
  \caption{\label{fig:hist_eps_gamma}%
    Distribution  of  $\cos\eps$ for  $  0  \leq \gamma  <  1.8^\circ$
    (blue),   $43.2^\circ  \leq   \gamma   <   45^\circ$  (red),   and
    $88.2^\circ\leq \gamma \leq 90^\circ$ (green).  
    The left-hands panels show  close pairs with normalized separation
    $0<p =  d/R_\text{vir,N} <0.66$,  and the right-hand  panels loose
    pairs with $0.85<p<1$. 
    The top and bottom rows respectively show the results at redshifts
    0 and 1.
  }
\end{figure}

Figure  \ref{fig:sketch_pro}  illustrates   the  distribution  of  the
positions and  the orientations  of the major  axis of  the neighbours
with respect to the target.  
We  divided  the sample  of  neighbours  into three  equal  subsamples
according to 
their separation normalized to the virial radius of the neighbour
$p = d/R_\text{vir,N}$, and show the 33\% closest pairs ($0<p<0.66$, blue)
and  33\% further apart pairs ($0.85<p<1$, red). 
Neighbours  are  randomly  drawn  at  angles  $\gamma$  following  the
acceptance-rejection  method: for  each  bin of  $\gamma$,  we drew  a
uniformly-distributed  random number  and compared  to the  normalized
distribution of $\gamma$.  
If  the random  number  is  smaller than  the  probability density  of
$\gamma$, we accept it and draw a neighbour at this $\gamma$.
Since $\gamma$ is defined between 0 and \SI{90}{\degree}, we populated
the 4 
quadrants  by  drawing four  independent  random  realisations of  the
positions. 
It clearly appears  that the neighbours are  preferentially located at
low latitudes  with respect to the  major axis of the  target ($\gamma
\simeq 0$ or \SI{180}{\degree}).  
At a given $\gamma$, we randomly picked a neighbour falling into this
bin, and placed it at its  normalized separation $p$ and orientate its
major axis  with an angle  $\eps$ from the  line connecting it  to the
target.  
As  expected  from  Fig.~\ref{fig:coseps},   the  major  axis  of  the
neighbour is  on average aligned  with the  line connecting it  to the
target.

This is  confirmed in Fig.  \ref{fig:hist_eps_gamma},  which shows the
distribution of $\cos\eps$ for fixed  ranges of gamma: $0\leq \gamma <
1.8^\circ$ (blue), 
$43.2^\circ\leq\gamma < 45^\circ$
(red),   and   $88.2^\circ\leq\gamma<90^\circ$  (green),   for   close
($0<p<0.66$, left) and 
distant ($0.85<p<1$, right) interactions, and  at redshifts 0 (top) and
1 (bottom).  
At all epochs and normalized distance  $p$, there is a strong alignment
signal in $\cos\eps$.
For $0.85\leq p < 1$ (loose  pairs), the distribution of $\cos\eps$ is
independent of 
the position angle $\gamma$.
However,  for $0\leq  p <  0.66$  (close pairs),  the distribution  of
$\cos\eps$ depends on $\gamma$. 
For $\gamma  \simeq 45-90^\circ$,  the distribution  of $\cos\eps  $ is
similar to that of the loose pairs. 
However, at small $\gamma$, the alignment signal is stronger. 
This  shows that  the alignment  of  the major  axes is  given by  the
large-scale  structure,  since it  is  present  at  all $p$,  and  the
influence of the  neighbours increases this alignment  signal at small
separations.
{%
  Fig.~\ref{fig:hist_eps_gamma}  shows  that  the major  axis  of  the
  neighbour halo is  pointing toward the target halo in  both loose and
  close pair  cases. On  the other  hand, the  degree of  alignment is
  independent of the  relative angular position of  the neighbour halo
  in the  case of  loose pairs, but  depends on it  in the  close pair
  case.
  We interpret  this latter phenomenon due  to the tidal force  of the
  target halo on the close neighbour.
  We attribute the  alignment of loose pairs, which is
  independent  of the  internal property  of the  target halo,  to the
  tidal shear force set up by the large-sale mass distribution. 
}

\subsection{Spin alignments}

\begin{figure}
  \begin{center}
    \begin{tikzpicture}
      \filldraw[color=blue!60, fill = blue!5,thick]
      (0,0)ellipse(1.5 and 0.5); 
      \filldraw[color=red!60,       fill      =       red!5,thick,rotate
      around={30:(2,2)}] %
      (2,2) ellipse (1.5 and 0.5); 
      \draw [->, color=blue!50,thick] (0,0)--(0,2) ; 
      \draw [->, color=blue!50,thick] (0,0)--(2,2) ; 
      \draw [->, color=blue!50,thick] (2,2)--(2,4) ; 
      \draw [->, color=red!50,thick] (2,2)--(1,3.73205080757) ; 
      \node[anchor=east,color=blue] (T) at (0,0) {T};
      \node[anchor=east,color=red] (N) at (2,2) {N};
      \draw[->, color = red!50,thick](2,3) arc (90:120:1);
      \node[color=red,anchor=north east] (phi) at (2,3) {$\phi$}; 
      \node[anchor=south,color=blue]        (at)         at        (0,2)
      {$\mathbf{J}_\text{T}$}; 
      \draw[->, color = blue!50,thick](0.705,.705) arc (45:90:1);
      \node[color=blue] (alpha) at (.2,.8) {$\alpha$}; 
      \node[anchor=south,color=red]     (an)    at     (1,3.73205080757)
      {$\mathbf{J}_\text{N}$}; 
      \node[anchor=north,color=blue] (at) at (2,2) {$\mathbf{r}$}; 
    \end{tikzpicture} 
    \caption{\label{fig:def_aphi}  Definition of  the angles  $\alpha$
      between  the  spin of  the  target  $\vect{J}_\text{T}$ and  the
      direction of  the neighbour  $\vect{r}_\text{N}$, and  the angle
      $\phi$ between $\vect{J}_\text{T}$ and the spin of the neighbour
      $\vect{J}_\text{N}$.  
    }
  \end{center}
\end{figure}
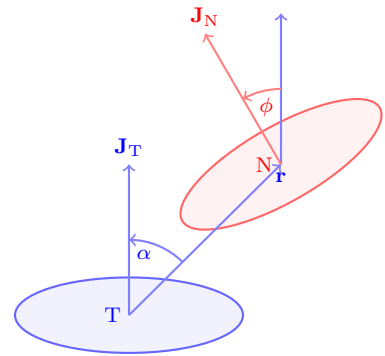

In this  section, we  study the  alignment of the  spin of  the target
haloes with respect to their neighbours.  
We define the angle $\alpha = (\vect{J}_\mathrm{T},\vect r_\text{N})$
between the spin of the target and the direction to the nearest neighbour, and
$\phi  = (\vect  J_\text{T},\vect J_\text{N})$  the angle  between the
spins of the target and neighbour  (see
Fig.~\ref{fig:def_aphi}).
Since spins have a direction, the  angles are considered between 0 and
\SI{180}{\degree}.

\begin{figure}
  \begin{center}
    \includegraphics[width=\columnwidth]{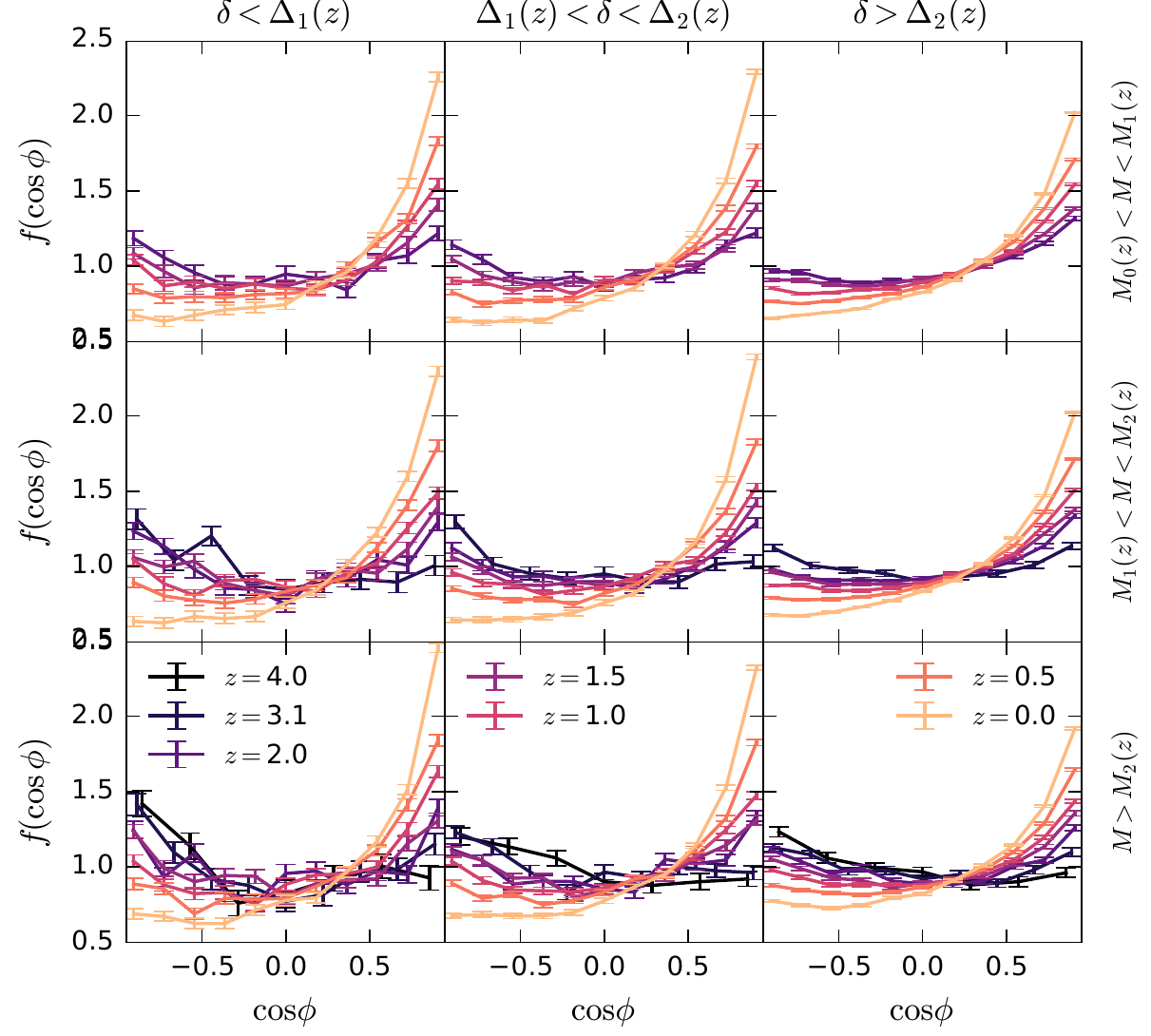}
    \caption{\label{fig:cosphi}Time evolution of the distribution of 
       $\cos\phi$,    where    $\phi    =    (\vect    {J_\mathrm{T}},
      \vect{J_\mathrm{N}  })$ is  the angle  between the  spin of  the
      target and that of the neighbour.  
      The  legend is the same  as Fig.~\ref{fig:cosgamma}.
    }  
  \end{center}
\end{figure}

\begin{figure}
  \includegraphics[width=\columnwidth]{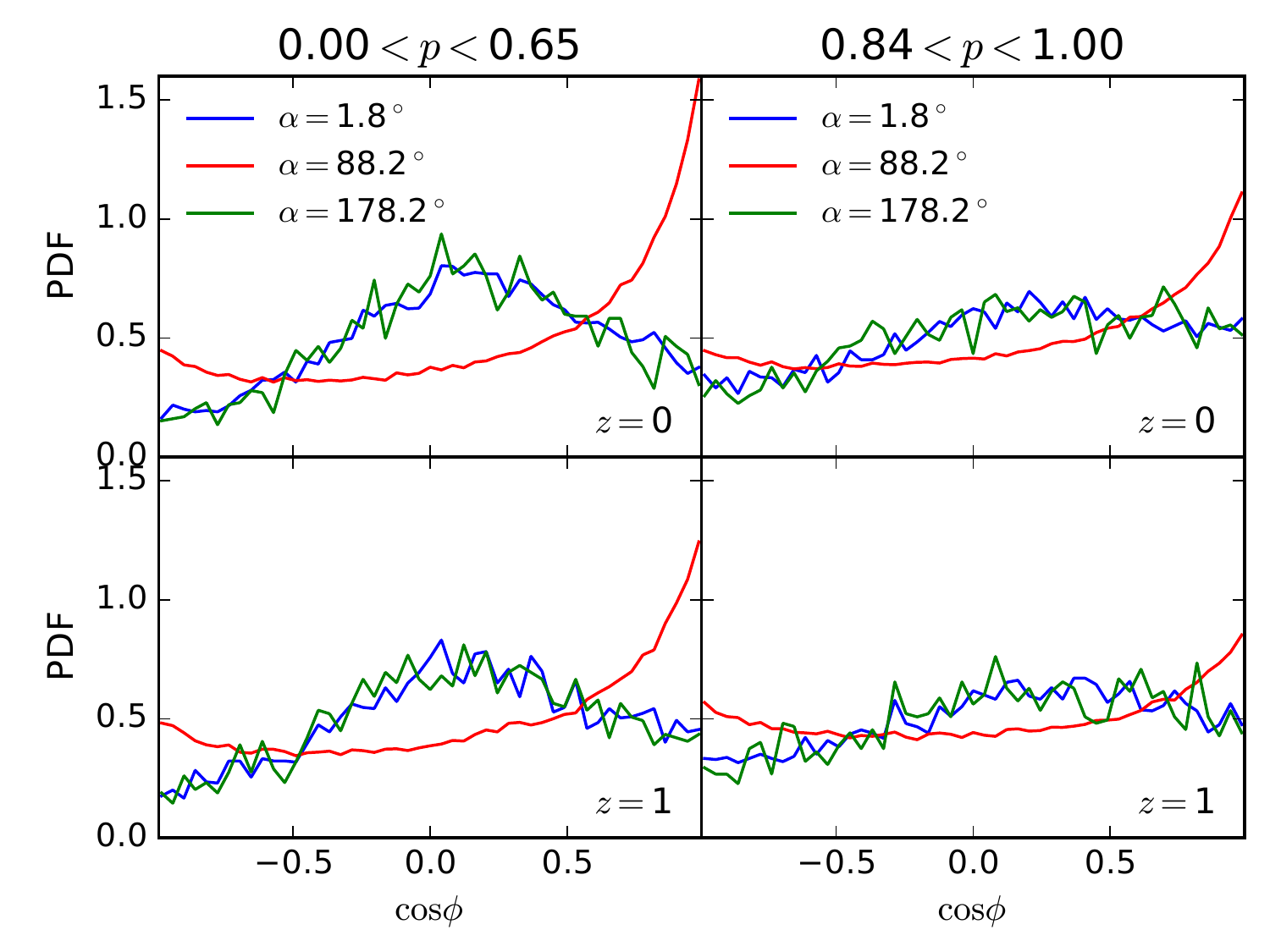}
  \caption{\label{fig:hist_phi_alpha}%
    Distribution of $\cos\phi$ for  $0\leq \alpha < 3.6^\circ$ (blue),
    $86.4^\circ  \leq\alpha< 90^\circ$  (red),  and $176.4^\circ  \leq
    \alpha \leq 180^\circ$ (green).  
    The left  panels are  for close  interaction with  normalized halo
    separation for the 33\% closest pairs
    $0<p = d/R_\text{vir,N} < 0.65$, and  the right panel for the 33\%
    furthest apart pairs ($0.84<p<1$).  
    The top row is for redshift 0 and the bottom row at redshift 1.
  }
\end{figure} 

Figure~\ref{fig:cosphi} shows  the alignment between the  spins of the
target and neighbour.
At  $z=4$,  there   is  a  slight  preference   for  an  anti-parallel
configuration  ($\cos\phi \simeq  -1$)  in all  environment, which  is
stronger for lower densities.  
At  $z=3.1$,  the  situation  is  similar  at  low-  and  intermediate
densities, but  starts to  change in  high-density regions,  where the
distribution  becomes  more  symmetric, and  the  parallel  ($\cos\phi
\simeq 1$) and anti-parallel ($\cos\phi \simeq -1$) configurations are
of the same order.  
At  $z=2$,  the parallel  configuration  is  dominant in  high-density
regions,  while  the situation  is  becoming  symmetric in  the  lower
density bins.
Eventually, at  all mass and  density, by $z=0$, the  alignment signal
for the parallel  configuration becomes stronger and  stronger and the
anti-parallel configuration is almost lost.
Interestingly,  the alignment  signal  is stronger  at lower  density,
where the initial configuration was mostly anti-parallel.
{
  Similar  results  to  those from  Fig.~\ref{fig:cospsi}  can  be
  achieved for $\cos\alpha$, we omitted them for the sake of clarity.
}

We  can illustrate the alignment of spin as  Fig.~\ref{fig:sketch_pro}. 
This time, we do not need  to restrict ourselves to prolate pairs, and
consider all interacting pairs.
However, since the alignment signal is  weaker for the spins, a figure
similar to Fig~\ref{fig:sketch_pro} is not as instructive as before.

\begin{figure}
  \begin{center}
    \begin{tikzpicture}
      \node[color=blue] (target) at (0,0) {T} ;
      \node[color=blue](neighbour) at (2,0) {};
      \node[color=blue] (N) at (neighbour.south east) {N};
      \draw [->,color=blue,thick] (target) -> (neighbour);
      \draw [color=blue,thin, dashed] (target) -> (3,0);
      \node[color=blue,anchor=north] (pos) at (1.5,0) {$\mathbf{r}$} ; 
      \node[color=blue](vel) at ( 3.29903811,  0.75) {};
      \node[color=blue](v) at (vel.east) {$\mathbf{v}$};
      \draw[->,color=blue,thick] (neighbour) -> (vel);
      \node(theta) at (3,0) {};
      \draw[->,color=blue] (theta) arc (0:30:1);
      \node[color=blue] at (theta.north east) {$\theta$};
      \draw [->,color=red,thick] (neighbour) -> (2,1.5);
      \node[color=red] (L) at (1.8,1.5) {$\vect {L}_\text{N}$};
      \node[color=red] (Jn) at (1.5, 0.8660254) {};
      \node[color=red] (vJn) at (Jn.west) {$\vect
        J_\text{N}$};
      \draw[->,color=red] (neighbour) -> (Jn);
      \node (beta) at (2,.75) {};
      \draw[->,color=red] (beta) arc (90:120:.75);
      \node[color=red] () at (beta.north west) {$\beta$}; 
      \draw [thin, draw=blue, fill=blue, fill opacity=0.1]
      (-1.5,-1) -- (0.5,1) -- (4.5,1) -- (2.5,-1) -- cycle;
      \draw [thin, draw=red, fill=red, fill opacity=0.1]
      (3,1)  --   (3,3)   --
      (1, 1) -- (1,-1) -- cycle; 
    \end{tikzpicture}
    \caption{%
      \label{fig:beta}Definition  of the  angles $\theta  = (\vect  r,
      \vect v)$ and $\beta  = (\vect {L}_\text{N}, \vect {J}_\text{N}$),
      where $\vect {L}_\text{N} = \vect r \times \vect v$.  
    }
  \end{center}
\end{figure}
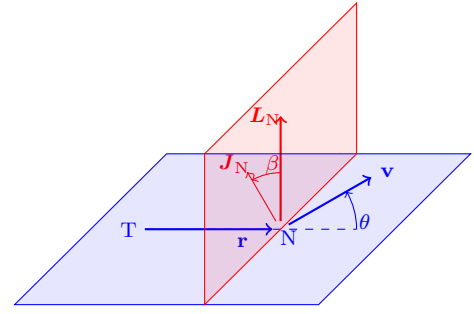

Instead, we show in  Fig.~\ref{fig:hist_phi_alpha} the distribution of
$\cos\phi$, the angle between the halo spins in three different ranges
of $\alpha$. 
We divide the sample of pairs into three equal subsamples according to
their normalized separation $p$. 
We find that the spin--spin alignment changes sensitively depending on
the relative angular position of the nighbour with respect to the spin
of the target.
For  neighbours  located  at  high latitudes  ($\alpha  \simeq  0$  or
$180^\circ$), the  spins tend  to be  orthogonal, while  at equatorial
latitudes  ($\alpha  \simeq  90^\circ$),   $\cos\phi$  has  a  bimodal
distribution peaking at $\cos\phi= \pm 1$, corresponding to parallel 
and anti-parallel cases.
In  all  cases,  the  distribution of  $\cos\phi$  is  skewed  towards
positive alignment ($\mean{\cos\phi}>0$).
We  saw  in   Fig.~\ref{fig:cospsi}  and~\ref{fig:cosgamma}  that  the
equatorial  configuration  is  preferred, which  explains  the  larger
fluctuations       at      $1.8$       and      $178.2^\circ$       in
Fig.~\ref{fig:hist_phi_alpha}.
The preference  for $\alpha\simeq 90^\circ$  is also responsible  for the
apparent bimodal distribution of $\cos\phi$ in Fig.~\ref{fig:cosphi}: 
neighbours located  in the  equatorial plane tend  to have  their spin
either  parallel or  anti-parallel to  the target  (red lines),  while
those  located in  polar directions  have their  spin orthogonal,  but
are  less common  and  thus  only weekly  affect  the distribution  of
$\cos\phi$.  
Similarly  to Fig.~\ref{fig:hist_eps_gamma},  the alignment  signal is
affected by  the normalized separation:  it is slightly weaker  in the
  loose pairs, while close  interactions tend to
have a stronger interaction signal. 
However, the effect of the distance  on the alignment is not as strong
as in Fig.~\ref{fig:hist_eps_gamma}.
{%
  We      checked     that      the     stronger      alignment     in
  Fig.~\ref{fig:hist_eps_gamma} 
  is not due to the the selection criteria for oblate haloes ($q<0.8$)
  by applying the same criterion in Fig.~\ref{fig:hist_phi_alpha}.
  The stronger alignment of the shapes  is thus a real effect, and can
  be understood as follows:
  The  initial  alignment  of  the  spins  and  the  shapes  with  the
  large-scale structure comes from the tidal field (TTT).
  However, as structure evolve non-linearly through merger events, the
  alignment  becomes  more complex:  merger  can  completely flip  the
  spins \citep{2002MNRAS.332..325P}, while the shapes are more robust.
  Therefore, it  is not surprising  that the alignment is  stronger in
  the case of the shapes.
}

\begin{figure}
  \centering
  \includegraphics[width=\columnwidth]{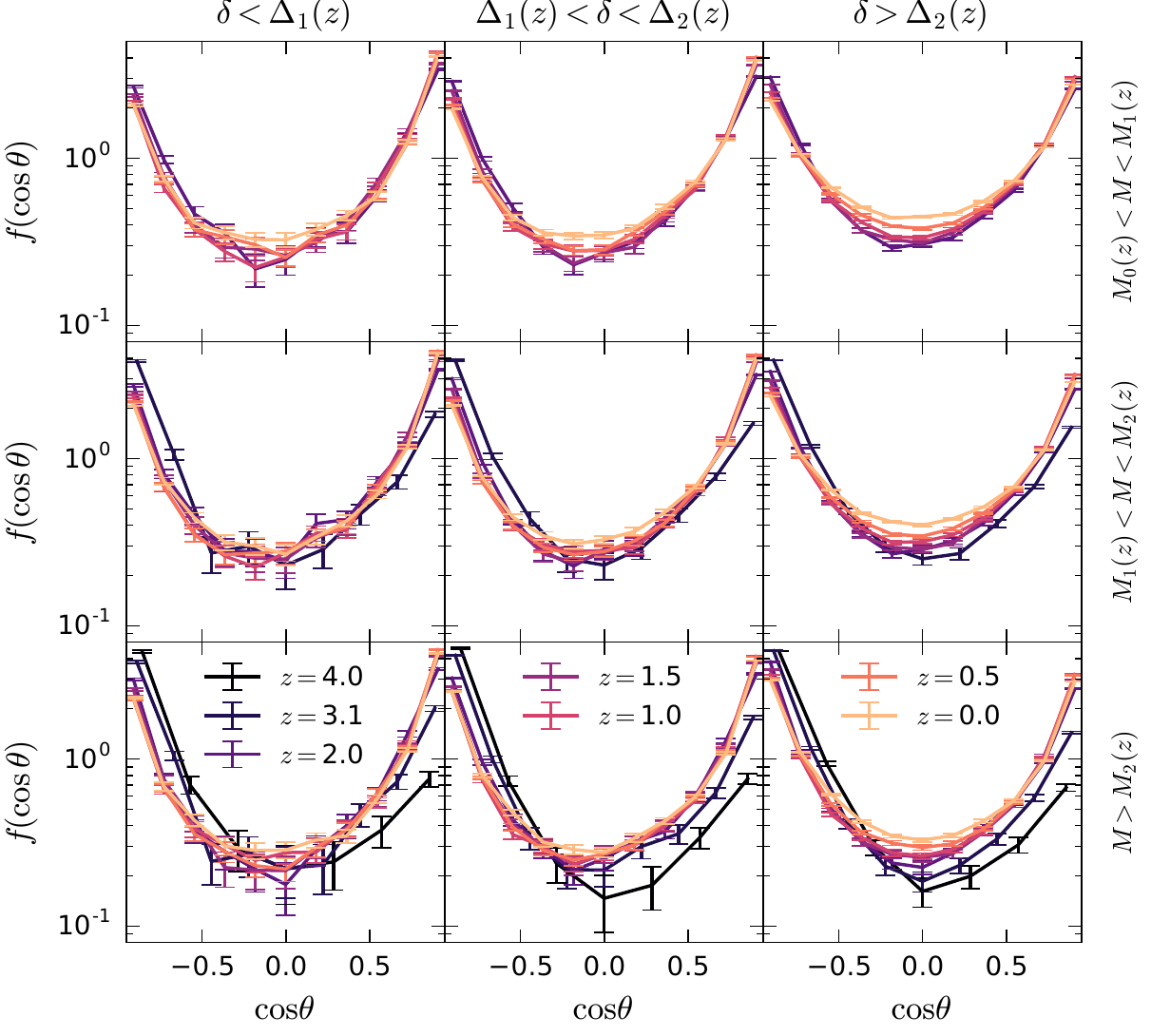}
  \caption{\label{fig:costheta}
    Distribution of $\cos \theta$, where $\theta = (\vect r, \vect v)$
    is  the angle  between  the  velocity and  the  position of  the
    neighbour in the frame of the target halo, same legend
      as Fig. \ref{fig:cosgamma}.  
  }
\end{figure}

\subsection{Orbits of the interactions}

\label{sec:angles} 
\subsubsection{Incident angle}

In this  section, we study the distribution of  the angle
$\theta = (\vect r, \vect v)$ between the position and velocity of the
neighbour in the frame of the target (see Fig.~\ref{fig:beta}).
A  value  of  $\cos\theta  \simeq(-)1$ means  that  the  satellite  is
radially  receding from  (approaching)  the  target, while  tangential
orbits have $\cos\theta \simeq 0$.  

Figure  \ref{fig:costheta}   shows  the   normalized  pair   count  of
$\cos\theta$ for the same ranges of  density and target mass as in the
previous section.  
At  all   masses,  densities,   and  redshifts,   there  is   a  clear
alignment signal: the orbits are preferentially radial.
At early  times, $z\geq  3$, there is  a peak at  $\cos \theta  = -1$,
corresponding to radially infalling orbits.
The  peak  decreases  with  decreasing  redshift,  and  by  $z=2$  the
distribution  of  angles becomes  more  symmetric  and the  number  of
radially receding orbits increases.
Meanwhile, the  contribution of orbits with  $\abs{\cos\theta} \ll 1$,
corresponding to tangential orbits, increases.
This  effect  was  already seen  in  \citetalias{2015MNRAS.451..527L},
where  we separated  the orbits  according  to their  position in  the
$(p,\xi)$   plane\footnote{In  \citetalias{2015MNRAS.451..527L},   the
  mass ratio  was denoted as $q$.  However here, $q$ was  already used
  for the oblateness.},  where $p$ is the pair separation  in units of
virial radius of the neighbour, and $\xi$ is the mass ratio.  
We showed  that the typical trajectory  of the orbit is  radial at the
first encounter, then becomes more random.

The  effects of  mass and  density on  the normalized  pair count  are
weak. 
At  low density,  the radially  receding orbits  start to  dominate at
$z=2$,  while  at  intermediate  densities, they  become  dominant  at
$z=1.5$.
In the  higher-density bin, they  become dominant at $z=1$,  except in
the higher mass bin where radially receding orbits are the same order as
radial infalling.

\begin{figure}
  \centering
  \includegraphics[width=\columnwidth]{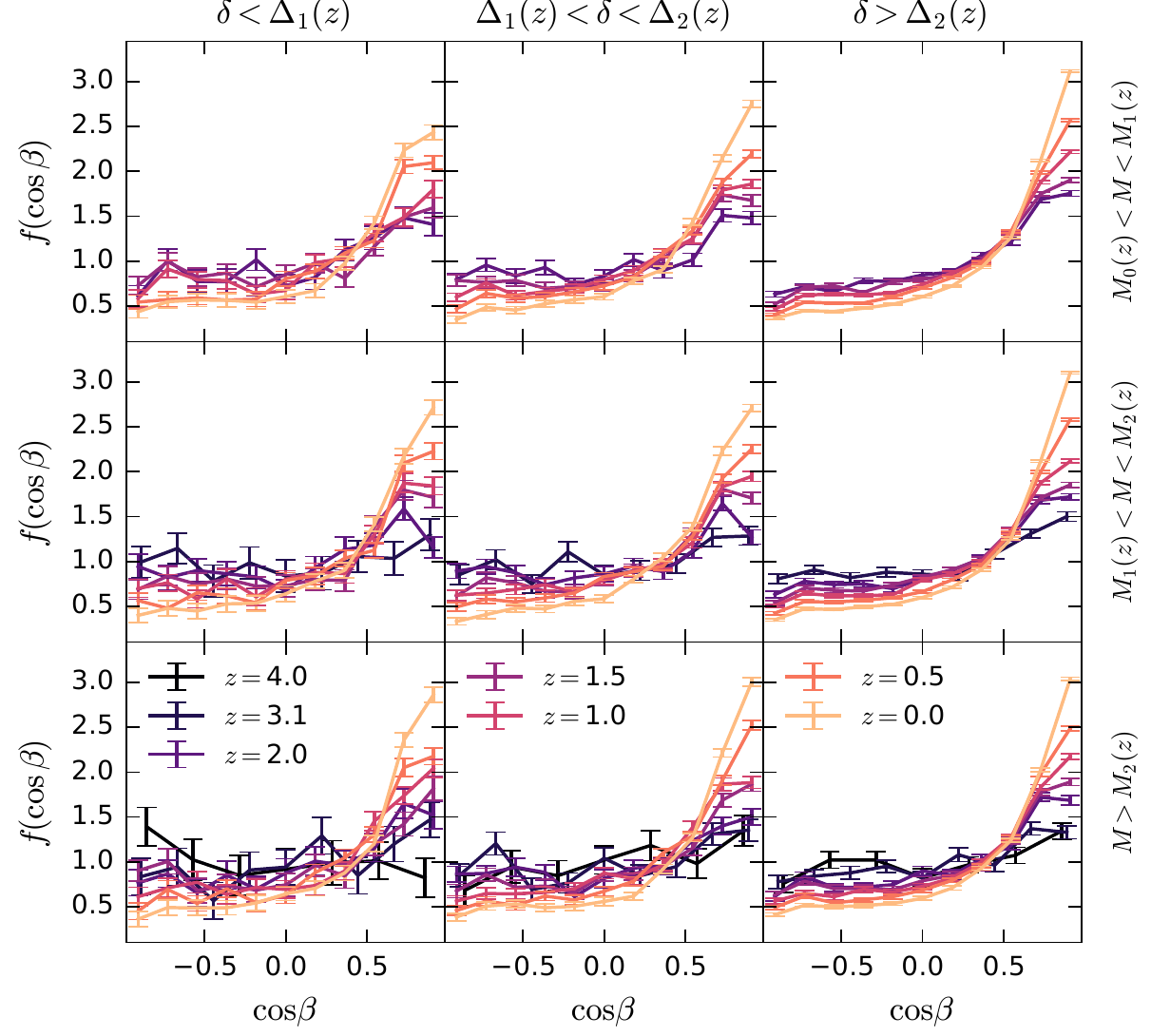}
  \caption{\label{fig:cosbeta}
    Distribution   of   $\cos   \beta$,    where   $\beta   =   (\vect
    {L}_\text{N},\vect  J_\text{N})$  is  the  angle  between  orbital
    angular momentum of the target and $\vect J_\text{N}$ its spin. 
  }
\end{figure}
 
\subsubsection{Prograde versus retrograde encounters}

Finally, we  looked at the  angle $\beta = (\vect  {L}_\text{N}, \vect
J_\text{N})$ between the spin of the neighbour and its orbital angular
momentum $\vect {L}_\text{N} = \vect r \times \vect v$.  
A  value of  $\cos\beta  =  (-)1$ means  that  the  orbit is  prograde
(retrograde),  while  $\cos\beta =  0$  either  means a  radial  orbit
($\vect {L}_\text{N} = 0$) or an orbit in the plane of rotation of the
neighbour ($\vect {L}_\text{N} // \vect J_{\text{N}}$).  
Therefore, we restricted ourselves  to cases where $\abs{\cos\theta} <
0.5$, ensuring a non-radial orbit.  

Figure  \ref{fig:cosbeta}   shows  the   normalized  pair   count  for
$\cos\beta$.  
At  $z=4$,  the  alignment  signal is  consistent  with  no  alignment
(i.e. random orbits) for low- and
intermediate  densities, while  at high-densities,  there is  a slight
excess of  prograde orbits  ($\cos\beta \simeq  1$), and  depletion of
retrograde orbits ($\cos\beta \simeq -1$). 
At  lower  redshifts, the  excess  of  prograde encounters  starts  to
develop at lower densities and at all masses too. 
Overall, the alignment signal is  the strongest at high density, while
the effects of mass in the range probed by the constant-number density
bins is very weak.

\subsection{Dependence on mass and density}

\label{sec:align_md}

\begin{figure}
  \begin{center}
    \includegraphics[width=\columnwidth]{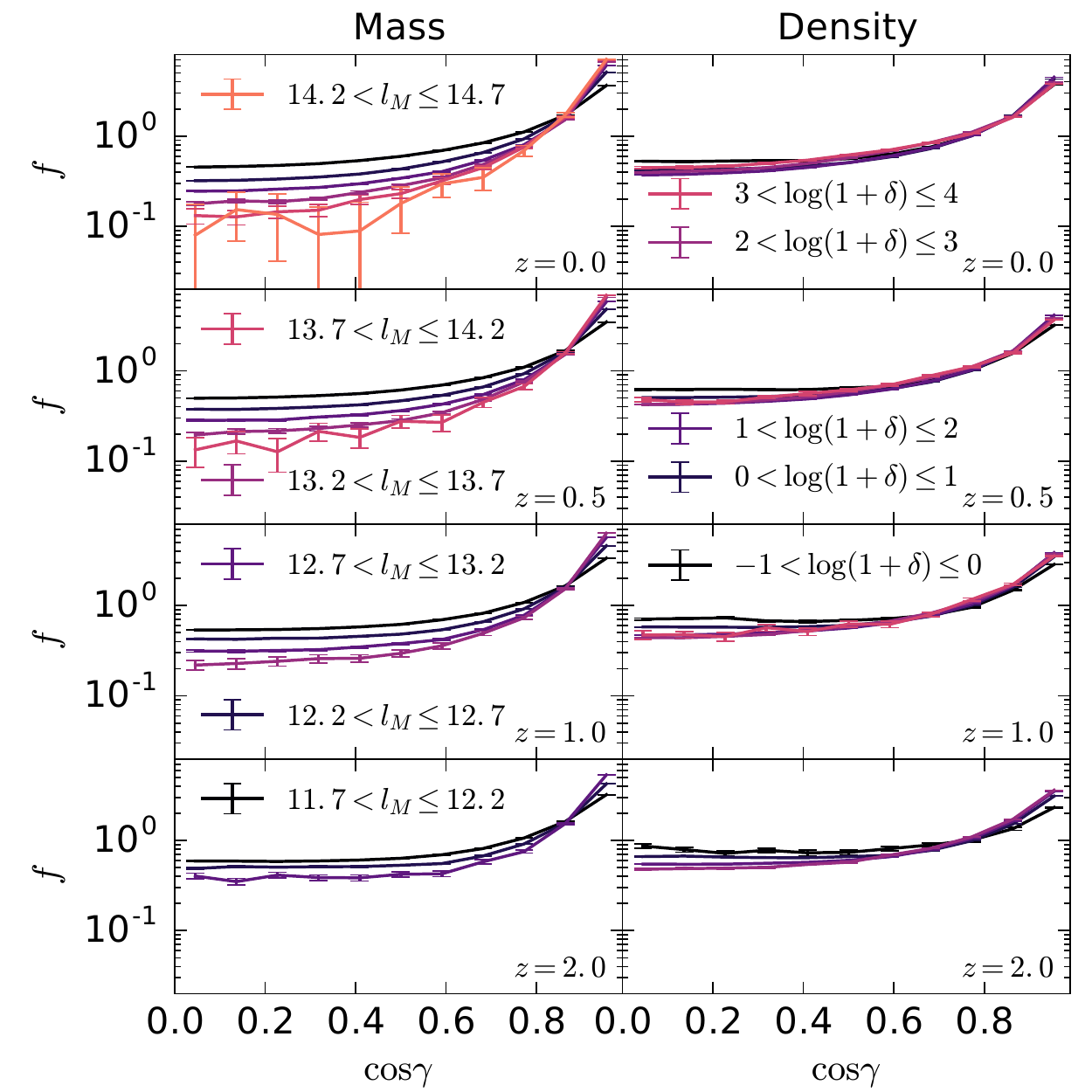}
    \caption{\label{fig:gamma_md}
      Effect of mass and density  on the distribution of $\cos\gamma$,
      where $\gamma = (\vect a_\text{T},\vect{r})$ is the direction of
      the nearest  neighbour halo  relative to the  major axis  of the
      target. 
      Left: effect of  mass, using six bins with $\Delta  l_m = 0.5$,
      and $l_M=\log_{10}(h M / M_\odot)$.
      Right:  effect of  the large-scale  density, using  five density
      bins with $\Delta \log_{10}(1+\delta) = 1$. 
    }  
  \end{center}
\end{figure} 

\begin{figure}
  \begin{center}
    \includegraphics[width=\columnwidth]{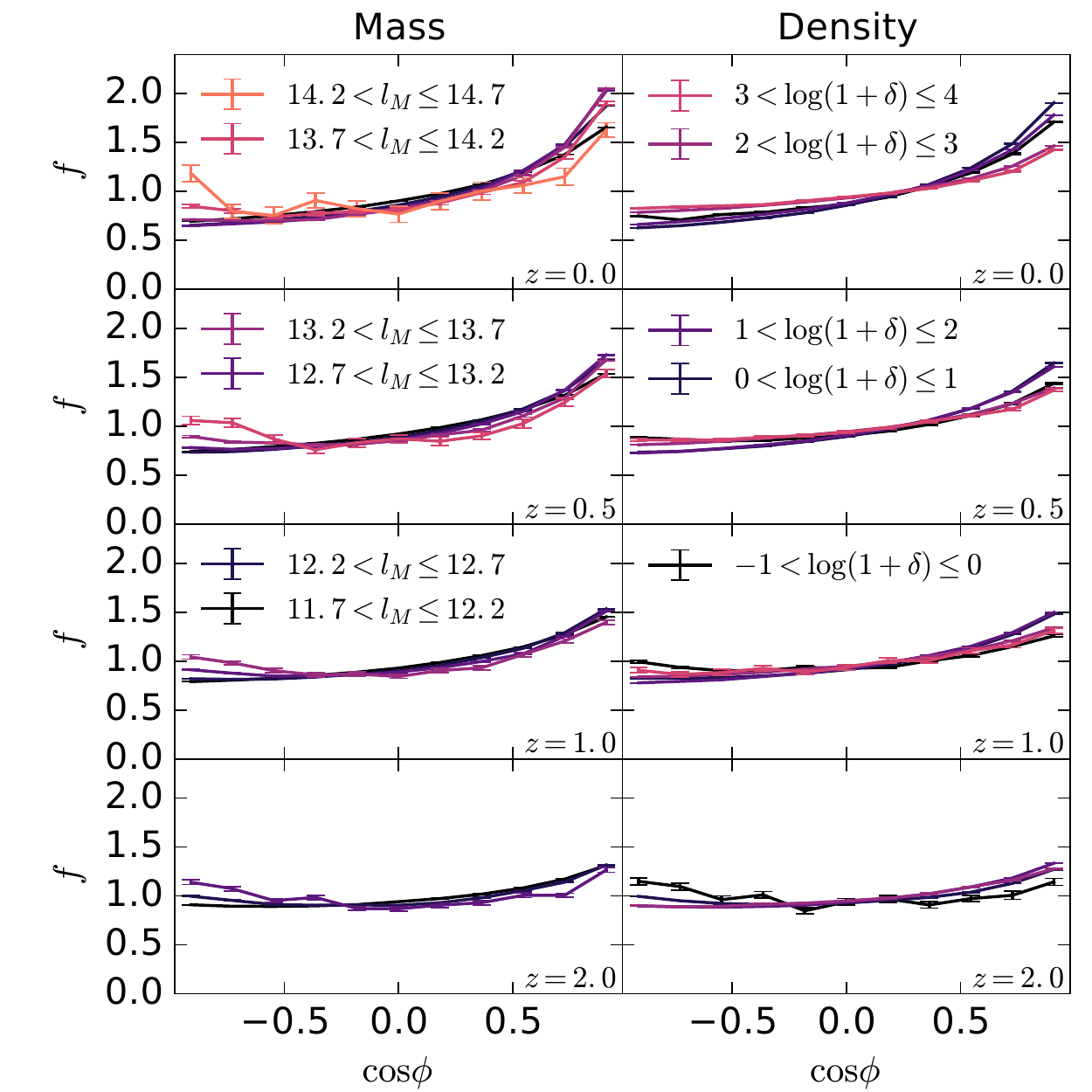}
    \caption{\label{fig:phi_md}
      Effect of  mass and  density on  the alignment  of the  spins of
      interacting halos,  as shown by the  distribution of $\cos\phi$,
      where $\phi = (\vect J_\text{T},\vect{J}_\text{N})$. 
      Same legend as Fig.~\ref{fig:gamma_md}
    }  
  \end{center}
\end{figure} 

\begin{figure}
  \begin{center}
    \includegraphics[width=\columnwidth]{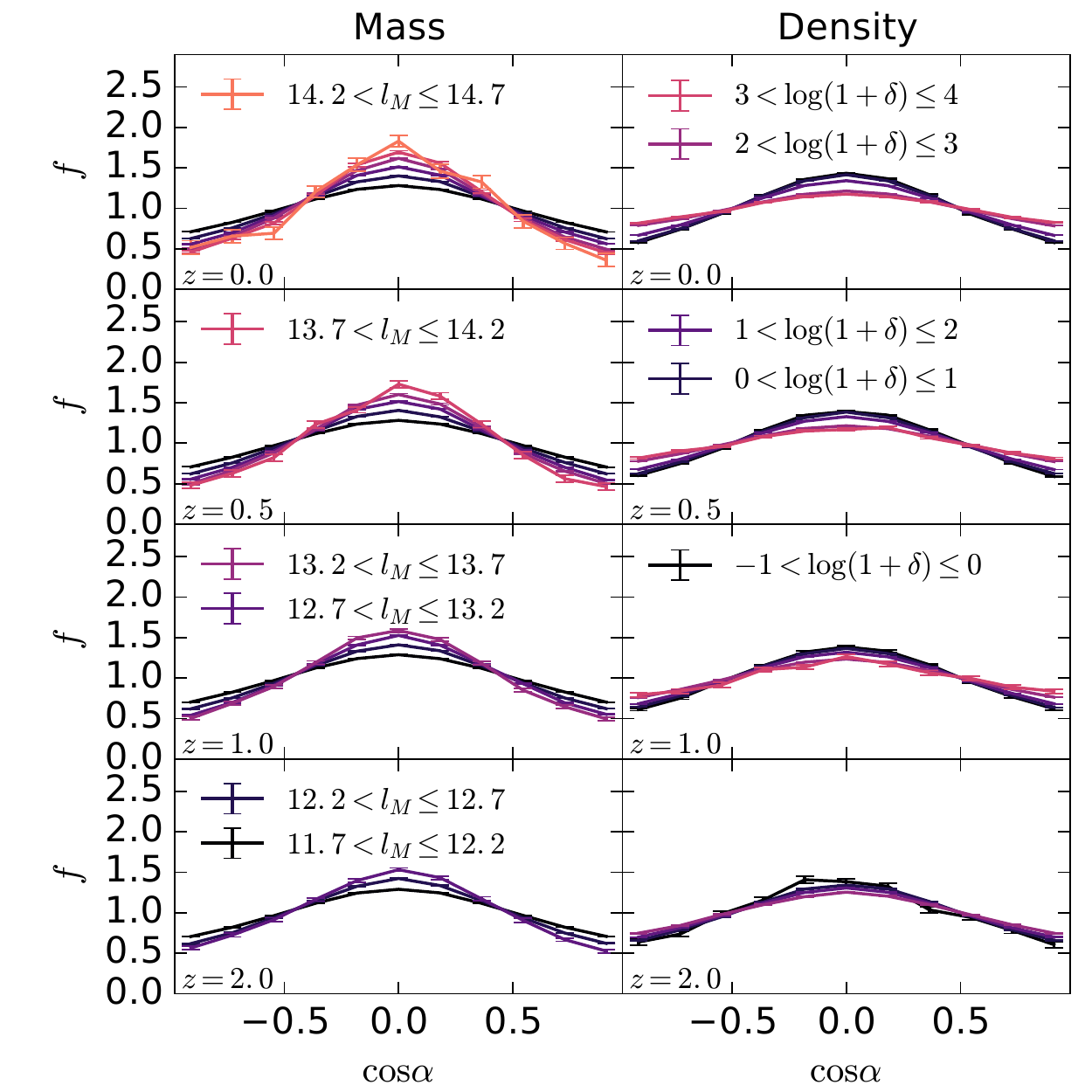}
    \caption{\label{fig:alpha_md}
      Effect of mass and density on  the alignment of the direction of
      the  neighbour and  the  spin of  the target,  as  shown by  the
      distribution of
      $\cos\alpha$,
      where $\alpha = (\vect J_\text{T},\vect r)$. 
      Same legend as Fig.~\ref{fig:gamma_md}
    }  
  \end{center}
\end{figure} 
 
In this section,  we study  the dependence of  the alignment on mass
and density.
We focus on the angles $\gamma = (\vect {a}_\text{T}, \vect{r})$, and
$\phi = (\vect J_\text{T},\vect{J}_\text{N})$.
In the previous section, we kept  a constant number density of targets
in order to study the redshift evolution, which limited the sample at
low redshift.
In this  section, we  take advantage  of the whole  range of  mass and
density.
We focus on the alignment between  the position angle of the neighbour
and the orientation (major axis) of the target.

Fig.~\ref{fig:gamma_md}  and~\ref{fig:phi_md} respectively  show the
normalized  pair   counts,  as  defined   in  \S~\ref{sec:count},for
\abs{\cos\gamma} and $\cos\phi$ in 6  bins of mass (left column)
and  5  bins of  density  (right-hand  panels), equally  spaced  in
log-space.
Both angles show a strong alignment.
As for  $\cos\gamma$, the alignment  increase with redshift  and mass,
but the dependence on the large-scale density is very weak, especially
at high-redshift. 
$\cos\phi$  shows a  more complex  alignment, where  the alignment  is
weaker at low (dark) and high  (bright) mass and density, and stronger
at intermediate mass and densities. 
The maximum is reached for $1<1+\delta<10$ at every redshift, 
The lower alignment in low density is consistent with what is seen for
other quantities, and presumably due to weaker tidal fields.
At high density, multiple interactions with neighbours will flip the
spins  and weaken  the alignment,  while  this does  not affect  the
shapes as much.
The mass dependency is also weaker than in the shape alignment. 
Fig.~\ref{fig:alpha_md} shows  the alignment of the  position angle of
the neighbour with the spin of the target
($\alpha = (\vect J_\text{T}, \vect r)$). 
As expected  from Figs.  \ref{fig:cosgamma} and  \ref{fig:coseps}, the
neighbour is  preferentially located  in the  plane orthogonal  to the
spin of the target.  
The  anti-alignment signal  monotonically  increases  with the  target
mass, and decreases with the density.
This is  consistent with the findings  of \citet{2015arXiv151200400W},
where the authors study the evolution of the angle
$\theta_1 = (\vect c_\text{T}, \vect  r)$ between the direction of the
satellite and the minor axis of the central galaxy in the Horzizon-AGN
hydrodynamical simulation.
They  found  that as  the  satellite  enters  the virial  radius,  its
direction becomes anti-aligned with the minor axis of the central. 
This is interesting as it is directly observable.

The dependence on  mass and large-scale density is now  more clear, as
it  was  previously  limited  by  the  rather  small  dynamical  range
resulting from  the choice of a  constant number density of  target.
Using the whole  range of masses available from  the simulation enables
us to better  study the effects of mass, and  the apparently weak mass
dependency in the  previous section appears to be caused  by the small
range of masses probed.
However, in the  previous section, the use of  constant number density
enabled  us  to study  the  redshift  evolution  of the  alignment  by
following haloes that are statistically the same.
These two approaches are therefore complementary.


\section{Summary and Discussion}
\label{sec:ccl}

We used  the Horizon Run  4 cosmological $N$-body simulation  to study
the alignments  of interacting haloes  as a  function of the  mass and
large-scale density, pair separation, as well as their time evolution.
The  very large  volume of  the  simulation, combined  with its  large
number of particles, enabled us  to study the environmental effects on
interactions. 
Our main findings are as follow.

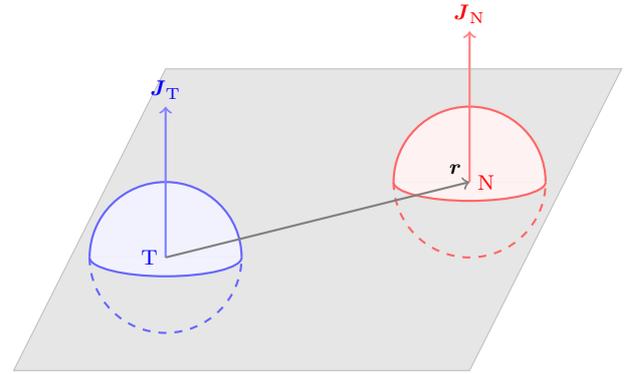
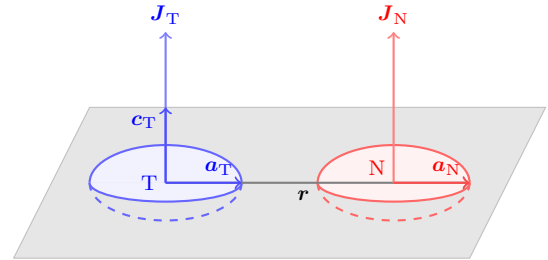
\begin{figure}
  \begin{center}
    \begin{subfigure}{\columnwidth}
      \begin{tikzpicture}      
        \filldraw[color=gray!50,fill=gray!20](-1,-1.5)--(1,2.5)--(7,2.5)--
        (5,-1.5) -- (-1,-1.5); 
        \filldraw[color=blue!60, fill = blue!5,thick] (2,0) arc (0:180:1); 
        \filldraw[dashed,color=blue!60,  fill =  gray!20,thick](0,0) arc
        (180:360:1);       
        \filldraw[color=blue!60,   fill   =  blue!5,thick]   (0,0)   arc
        (180:360:1 and .25); 
        \filldraw[color=red!60, fill = red!5,thick] (6,1) arc (0:180:1); 
        \filldraw[dashed,color=red!60, fill = gray!20,thick ] %
        (4,1) arc (180:360:1); 
        \filldraw[color=red!60, fill = red!5,thick ] %
        (4,1) arc (180:360:1 and .25); 
        \draw [->, color=blue!50,thick] (1,0)--(1,2) ; 
        \draw [->, color=black!50,thick] (1,0)--(5,1) ; 
        \draw [->, color=red!50,thick] (5,1)--(5,3) ; 
        \node[anchor=east,color=blue] (T) at (1,0) {T};
        \node[anchor=west,color=red] (N) at (5,1) {N};      
        \node[anchor=south,color=blue]        (at)         at        (1,2)
             {$\vect{J}_\text{T}$}; 
        \node[anchor=south,color=red]     (an)    at     (5,3)
             {$\vect{J}_\text{N}$}; 
        \node[anchor=south east] (at) at (5,1) {$\vect{r}$}; 
      \end{tikzpicture}
      \caption{\label{fig:sum_j}Alignment of spins in the plane of interaction}
    \end{subfigure}
    \begin{subfigure}{\columnwidth}
      \begin{tikzpicture}
        \filldraw[color=gray!50,fill=gray!20]
        (-1,-1)--(0,1) -- (6,1) -- (5,-1) -- (-1,-1);
        \filldraw[color=blue!60,  fill   =  blue!5,thick]   (2,0)  arc
        (0:180:1. and 0.5); 
        \filldraw[dashed,color=blue!60,  fill  = gray!20,thick]  (0,0)
        arc (180:360:1 and 0.5); 
        \filldraw[color=blue!60,  fill  = blue!5,thick]  (0,0)
        arc (180:360:1. and 0.25); 
        \filldraw[color=red!60,   fill   =  red!5,thick]   (5,0)   arc
        (0:180:1 and 0.5); 
        \filldraw[dashed,color=red!60, fill = gray!20,thick] (3,0) arc
        (180:360:1 and 0.5); 
        \filldraw[color=red!60,  fill   =  red!5,thick]   (3,0)  arc
        (180:360:1 and 0.25); 
        \draw [->, color=blue!50,thick] (1,0)--(1,2) ; 
        \draw [->, color=black!50,thick] (1,0)--(5,0) ; 
        \draw [->, color=red!50,thick] (4,0)--(4,2) ;
        \draw[->,color=blue!70,thick] (1,0) -- (2,0);
        \node[color=blue,anchor=south east] (at) at (2,0) {$\vect{a}_\text{T}$};
        \draw[->,color=blue!70,thick] (1,0) -- (1,1);
        \node[color=blue,anchor=north east] (ct) at (1,1) {$\vect{c}_\text{T}$};
        \draw[->,color=red!70,thick] (4,0) -- (5,0);
        \node[color=red,anchor=south east] (an) at (5,0) {$\vect{a}_\text{N}$};
        \node[anchor=east,color=blue] (T) at (1,0) {T};
        \node[anchor=south east,color=red] (N) at (4,0) {N};
        \node[anchor=south,color=blue]       (Jt)       at       (1,2)
             {$\vect{J}_\text{T}$}; 
        \node[anchor=south,color=red]     (Jn)    at     (4,2) {$\vect{J}_\text{N}$}; 
        \node[anchor=north east,color=black] (rr) at (3,0) {$\vect{r}$}; 
      \end{tikzpicture}
      \caption{\label{fig:sum_a}Alignment  of the  major  axes in  the
        direction of the interaction}
    \end{subfigure}
    \caption{\label{fig:summary}Summary: alignments  of the  spins and
      of the major axes} 
  \end{center}
\end{figure}
  
\begin{itemize}
\item
  Interacting  targets  have a  significantly  lower  spin and  higher
  oblateness and sphericity parameters than all targets.
  The spin of interacting haloes decreases with increasing redshift up
  to  $z\simeq  2$,   with  a  faster  decrease   at  lower  densities
  (Fig.~\ref{fig:spin}).  
  Meanwhile, their prolateness and sphericity decrease with decreasing
  redshift and separation.  
  The spin, sphericity, and oblateness  increase with density, and are
  essentially   constant   with   mass   (Fig.~\ref{fig:shape_all}).
\item Interactions  preferentially occur in  the plane of  rotation of
  the  target haloes  (the  spins  of targets  are  orthogonal to  the
  directions   of  the   neighbours,  figs.    \ref{fig:cosgamma}  and
  \ref{fig:cospsi}, and in  the direction of the major  axis of oblate
  haloes (Fig.~\ref{fig:cosgamma}).  
  However,  neighbours located  in the  direction of  the spin  of the
  target have their spin orthogonal (Fig.~\ref{fig:hist_phi_alpha}).
\item
  The alignments  of the position and  main axis of the  neighbour with
  the    main   axis    of   the    target   (Figs.~\ref{fig:cosgamma}
  and~\ref{fig:coseps})  are   stronger  than   those  of   the  spins
  (Figs.~\ref{fig:cosphi}.
  This can be understood by the  fact that mergers can flip the spins,
  while the shape are more robust.
\item
  We confirmed previous findings that the major axes of pairs are also
  aligned, and  the spins of targets  are aligned with the  minor axes
  \citep[][]{2006MNRAS.369.1293Y, 2007ApJ...655L...5A}.  
\item Interacting  pairs initially  have anti-parallel spins,  but the
  spins become parallel as time proceeds.  
  The alignment signal is stronger at lower densities.  
\item  The  orbits  are  preferentially radial,  initially  with  more
  approaching  orbits.  The  signal  become weaker  as  orbits  become
  randomized. 
\item  Pairs  with non-radial  trajectories  initially  have a  random
  alignment  between  the  spin  of  the  neighbour  and  its  orbital
  momentum, and the angles becomes more  and more aligned as time goes
  on (Fig.~\ref{fig:cosbeta}).
\item The signal of the alignment  of the neighbour with major axis is
  stronger for  more massive  haloes, but only  weakly depends  on the
  large-scale density (Fig.~\ref{fig:gamma_md}).
  As for  the spin-spin alignment (Fig.~\ref{fig:phi_md}),  the signal
  is stronger for intermediate  densities ($1<1+\delta <10$), since at
  higher densities, haloes experience many interactions which can flip
  the spins.
  The anti-alignment  of the position  angle of the neighbour  and the
  spin  of   the  target   (Fig.~\ref{fig:alpha_md})  can   be  tested
  observationally. 
\end{itemize}

We now have a clear picture of how haloes interact, as summarised in
Fig.~\ref{fig:summary}.
Starting   with   preferentially   anti-parallel  or   parallel   spin
orientations, on a radial orbit,  haloes tend to have their neighbours
aligned with their major axis, and orthogonal to their spin.
The alignments  become stronger and  stronger in time, while  the spin
anti-parallel configuration disappears.

The  alignments  between  the  major  axes  and  the  pair  separation
($\gamma$ and $\eps$) in Figs.~\ref{fig:cosgamma} and \ref{fig:coseps}
are in good qualitative agreement with observational work. 
For  instance, \citet{2009ApJ...703..951W}  studied  the alignment  of
$\theta_2$, the  angle between  the projected major  axis of  the host
group and  the line  connecting the host  and neighbour  groups, which
corresponds to our definition of $\gamma$.  
Note however that in their case, the separation to the neighbour group
may be larger than one virial radius away.
They found  a strong alignment  signal, with a preference  for $\theta
\simeq 0$, corresponding to $\cos\gamma \simeq 1$ in our case.
Moreover, when they restrict the sample to close pairs
($<3 R_\text{vir,N}$), the alignment signal is stronger, which is also
seen in our theoretical work.

Interestingly,   \citet{2009ApJ...703..951W}   did  not   detect   any
alignment signal for  $\theta_4$, which corresponds to  $\eps$ in this
work, even at smaller separations. 
Our work are in contradiction with  these results, since we do find an
alignment  signal, albeit  slightly  weaker, for  $\eps$  as shown  in
Figs.~\ref{fig:coseps} and ~\ref{fig:sketch_pro}.

Moreover,   the   spin--spin    and   spin--position   alignments   in
Fig.~\ref{fig:cosphi} is  in good agreement with  previous theoretical
studies, such as \citet{2012MNRAS.427.3320C},  who found that low-mass
haloes have a spin parallel to their filaments, while massive ones are
orthogonal.
Even  though we  did not  identify the  filaments in  this study,  the
direction of the nearest  neighbour may be used as a  proxy to that of
the filament \citep{2015A&A...576L...5T}.
We see  that the anti-alignment  between spin and  neighbour direction
becomes   stronger   as   time   passes   (see   Fig.~\ref{fig:coseps}
and~\ref{fig:cosgamma}.  
However, the masses of each bin are not fixed in time, but also grow. 
The  strong anti-alignment  at low-redshift,  corresponding to  higher
masses, can be interpreted as massive haloes with a spin orthogonal to
the filament, while the relatively  weaker alignment at high redshift,
corresponding to  lower halo  masses, may  be due  to the  cases where
low-mass  haloes have  their spin  in the  direction of  the filament,
pointing toward the neighbour.
This   qualitatively   agrees   with  the   mass-dependence   of   the
spin-filament alignment found in \citet{2012MNRAS.427.3320C}.

In a next step, we plan to study in more details the alignment of halo
pairs with the  large-scale structure and the cosmic web  by using the
Hessian of  the smooth density  field to characterise  the environment
\citep{2007MNRAS.381...41H, 2009MNRAS.396.1815F, 2013ApJ...762...72T}.

However,   this  study   was  performed   using  a   dark-matter  only
simulation. 
The  observed  shapes  of  galaxies come  from  the  baryonic  (stars)
component, which needs to be modelled. 
\citet{2010MNRAS.405..274H} used cosmological hydrodynamic simulations
to study the alignment of the galaxy  and halo spins, and only found a
weak alignment. 
In addition, it is interesting to see the effects of interaction on the
morphology of galaxies, for instance  to understand the effects of the
hot  gas   halo  on  morphology  transformation   during  interactions
\citep{2013JKAS...46...33K, 2013JKAS...46....1H, 2015ApJ...805..131H}.
It  is thus  crucial to  include hydrodynamics  in our  simulations to
extend our analysis in comparison with observations.
Direct comparison to previous observational work is difficult, because
different definition of interaction are adopted.
We plan  to use  existing observational data  for comparison  with our
predictions using  definition of  galaxy interactions  compatible with
ours.

Moreover, the existence of an alignment  at redshifts as high as $z=4$
suggests an origin in the initial density field.  
We will investigate this issue in a separate paper.

\appendix

\section{Effect of the weighting in the inertia matrix}
\label{sec:weight}
\begin{figure}
  \centering
  \begin{subfigure}[b]{.48\textwidth}
    \includegraphics[width=\columnwidth]{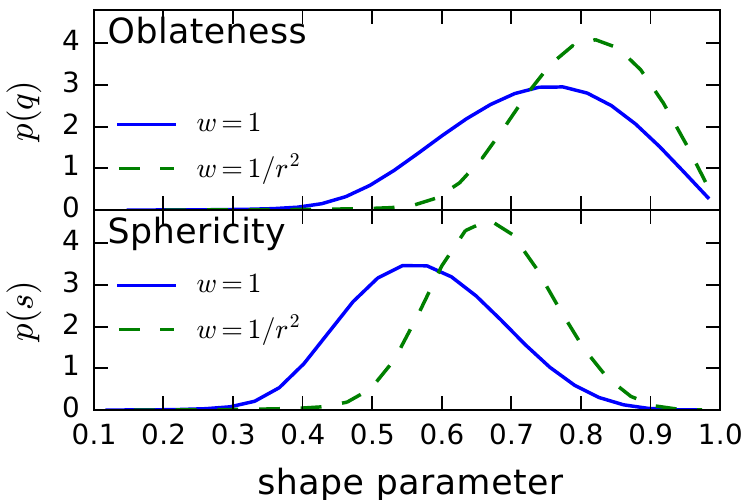}
    \caption{\label{fig:qsall_weight}%
      Distribution of the oblateness  (top) and sphericity (top)
      of target  haloes for  our two  assignment schemes  in the
      inertia  tensor: $w=1$  (solid  blue  line) and  $w=1/r^2$
      (green dashed line).
    }
  \end{subfigure}  
  \begin{subfigure}[b]{.48\textwidth}
    \includegraphics[width=\columnwidth]{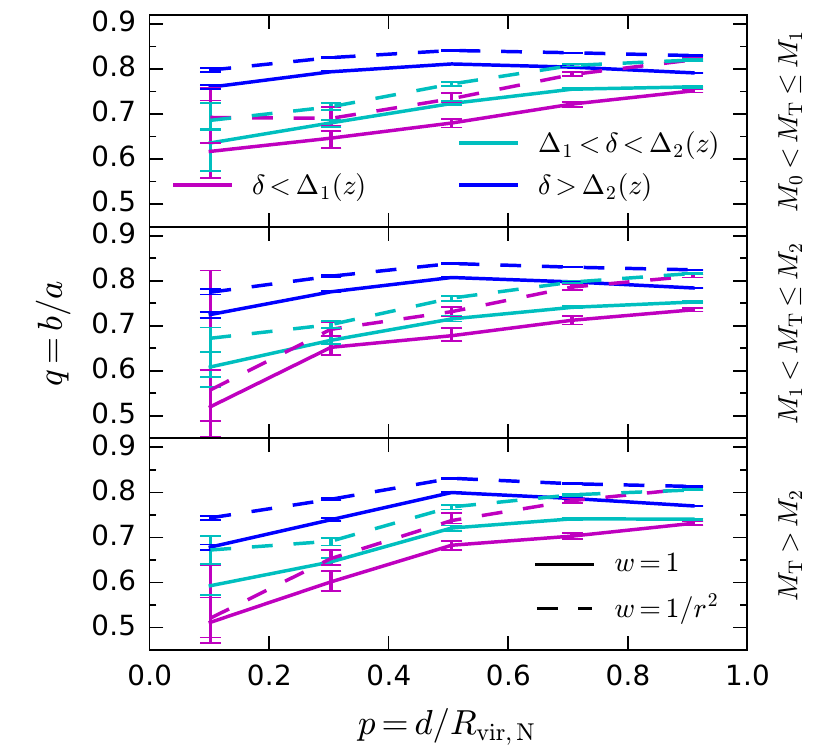}
    \caption{\label{fig:shapeq_p_weight}%
      Effect of the  weighting scheme of the  inertia tensor on
      the oblateness  $q$ as a  function of the  normalized pair
      separation $p$ at $z=0$.
    }
  \end{subfigure}
  \begin{subfigure}[b]{.48\textwidth}
    \includegraphics[width=\columnwidth]{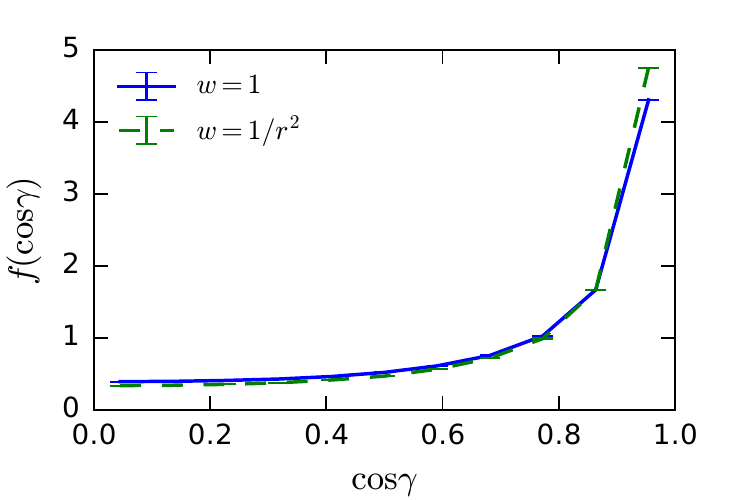}
    \caption{\label{fig:cosphi_weight}%
      Effect  of the  choice of  the weighting  of the  inertia
      tensor on  the alignment of  the major axis of  the target
      and the direction of the neighbour at $z=0$.
    }
  \end{subfigure}
  \caption{\label{fig:weight}%
    Effect of the weighting scheme  of the inertia tensor on the
    distribution  of  the  shape   parameters  (a)  and  on  the
    alignment  of  the major  axis  with  the direction  of  the
    neighbour (b).
  }
\end{figure}

The PSB algorithm truncates subhaloes at the tidal radius, which
can affect the shape measurement.
In  the  case  of  an  isothermal density  profile,  the  number  of
particles drops like $1/r^2$.
The  number of  particles in  spherical shells  with equal  bin size
is constant. 
Consequently, our method adopts an equal weighting in every shell.
When  applying a  different  weighing scheme,  such  as $1/r^2$,  to
eq.~\eqref{eq:shape}, innermost particles are favoured.
Of  course  our  method  is  susceptible  to  outer  boundary  noise
(numerical or varying with the subhalo definition).
But in  this study  we want  to study  the interactions  between the
nearest neighbours, which  may distort the halos shape  in the outer
boundary more seriously.
Therefore,   we    adopt   the   shape   tensor    as   defined   in
eq.~\eqref{eq:shape}.
In  this section, we  address this issue  by changing
the weighting scheme  in the calculation of  the inertia matrix
used to calculate the principal axes of the haloes.
A    more    comprehensive    study     can    be    found    in
\citet{2012MNRAS.420.3303B}.

We consider the generalised inertia tensor
\begin{align}
  \tilde{\tens{I}}_{ij}     &=     \sum_\alpha     w_\alpha     {\vect
    x_{\alpha,i}\vect x_{\alpha,j} }, 
\end{align}
where $\alpha$ is summed over all particles in the halo, and $i$
and $j$ are the considered directions.
Our choice corresponds to $w_\alpha=1$.
We compare it to another common choice, $w_\alpha=1/r_\alpha^2$.
In the following we will drop the $\alpha$ for clarity.

Fig.~\ref{fig:weight}(\subref{fig:qsall_weight})    shows    the
effect of weighting in the 
inertia tensor on  the distribution of the  shape parameters $q$
and $s$.
The  $w=1/r^2$  scheme  gives   more  weight  to  the  innermost
particles, so its measurement is  less sensitive to the external
part of the halo. 
The  $w=1/r^2$  yields  a  larger mean  sphericity  and  smaller
dispersion in both cases.
This shows that, even though  PSB truncates the subhaloes at the
tidal  radius,  the  sphericity  is  overall  dominated  by  the
innermost particles, therefore our results are weakly affected
by the effects  of the subhalo finder and the  truncation to the
tidal radius.
Panel~(\subref{fig:shapeq_p_weight}) shows the effect of the 
weighting scheme on the  evolution of the oblateness parameter
with the reduced pair separation $p$.
The colors are the same as in Fig.~\ref{fig:qshape_p}, but the
solid  line shows  the  $w=1$  case and  the  dashed line  the
$w=1/r^2$ weighting scheme.
While the actual oblateness is  higher for the $w=1/r^2$ case,
as    expected     from    the    distribution     in    panel
(\subref{fig:qsall_weight}), the trend is similar in all cases.
Fig.~\ref{fig:weight}(\subref{fig:cosphi_weight})    shows   the
effect of weighting the 
inertia matrix on the alignment of  the major axis of the target
and the direction of the  neighbour, defined as angle $\gamma$,
at $z=0$, and for all interacting targets.
Our  choice  of  $w=1$  is  shown in  blue,  solid  line,  while
$w=1/r^2$ is shown in green, dashed line. 
The alignment  signal is  slightly stronger  when $w=  1/r^2$ is
considered. 

Our choice  of $w=1$ is  thus a conservative definition  for the
inertia matrix regarding the alignment. 

\subsection*{Acknowledgements}
We thank KIAS Center for Advanced Computation for providing computing
resources.  

\bibliographystyle {mnras}

\bibliography{biblio}

\label{lastpage}
\end{document}